\documentclass[fleqn,usenatbib]{mnras}

\usepackage{natbib}
\usepackage{soul}

\usepackage[T1]{fontenc}
\usepackage{ae,aecompl}

\usepackage{outlines}
\usepackage{color}

\usepackage{graphicx}	
\usepackage{amsmath}	
\usepackage{amssymb}	
\usepackage{bm}	        
\usepackage{newtxtext,newtxmath}
\usepackage{hyperref}

\usepackage{ulem}

\usepackage{natbib}

\defcitealias{Mannucci2010}{M10}

\newcommand{\lcdm}{\Lambda\mathrm{CDM}}

\newcommand{\dd}{\mathrm{d}}

\newcommand{\ms}{M_*}
\newcommand{\msobs}{M_*^{\mathrm{obs}}}
\newcommand{\sfr}{\dot{M}_*}
\newcommand{\sfrobs}{\dot{M}_*^{\mathrm{obs}}}
\newcommand{\tausfh}{\tau_{\mathrm{sfh}}}
\newcommand{\mout}{\dot{M}_{\mathrm{out}}}
\newcommand{\zgas}{{Z}_{\mathrm{gas}}}
\newcommand{\zobs}{z_{\mathrm{obs}}}
\newcommand{\tobs}{t_{\mathrm{obs}}}
\newcommand{\etaceh}{{\eta}_{\mathrm{ceh}}}
\newcommand{\etakin}{{\eta}_{\mathrm{kin}}}
\newcommand{\mgas}{{M}_{\mathrm{gas}}}
\newcommand{\vmax}{V_{\mathrm{max}}}
\newcommand{\zo}{{Z}_{\mathrm{O}}}
\newcommand{\zoeq}{{Z}_{\mathrm{O}, \mathrm{eq}}}
\newcommand{\dzodt}{\mathrm{d}{Z}_{\mathrm{O}}/\mathrm{d}t}
\newcommand{\mocc}{{m}^{\mathrm{cc}}_{\mathrm{O}}}
\newcommand{\mo}{{M}_{\mathrm{O}}}
\newcommand{\modot}{\dot{M}_{\mathrm{O}}}

\newcommand{\mpc}{\mathrm{Mpc}}

\newcommand{\msol}{M_{\odot}}
\newcommand{\gyr}{\mathrm{G}yr}

\newcommand{\ssfr}{s{\mathrm{SFR}}}

\newcommand{\kms}{\mathrm{km}\,s^{-1}}

\newcommand\powexp{\texttt{powexp}}
\newcommand\eagle{\textsc{EAGLE}}

\newcommand{\rom}[1]{\uppercase\expandafter{\romannumeral #1\relax}}
\title[A Non-Equilibrium Chemical Evolution Model]{Constraints on galactic
outflows from the metallicity-stellar mass-SFR relation
of EAGLE simulation and SDSS galaxies}
\author[Lin \& Zu 2022]
{Yuanye Lin$^{1}$ and
Ying  Zu$^{1, 2, 3}$\thanks{E-mail: yingzu@sjtu.edu.cn}
\\ \\
$^{1}$Department of Astronomy, School of Physics and Astronomy, Shanghai Jiao Tong
University, Shanghai 200240, China\\
$^{2}$Shanghai Key Laboratory for Particle Physics and Cosmology, Shanghai Jiao Tong
University, Shanghai 200240, China\\
$^{3}$Key Laboratory for Particle Physics, Astrophysics and Cosmology,
Ministry of Education, Shanghai Jiao Tong University, Shanghai 200240,
China
}

\date{Accepted XXX. Received YYY; in original form ZZZ}

\pubyear{2023}

\begin{document}

\label{firstpage}
\pagerange{\pageref{firstpage}--\pageref{lastpage}}
\maketitle


\begin{abstract}
    Stellar feedback-driven outflows regulate the stellar formation and
    chemical enrichment of galaxies, yet the underlying dependence of mass
    outflow rate on galaxy properties remains largely unknown.  We develop
    a simple yet comprehensive non-equilibrium chemical evolution
    model~(NE-CEM) to constrain the mass-loading factor $\eta$ of outflows
    using the metallicity-stellar mass-SFR relation observed by SDSS at
    $z{=}0$.  Our NE-CEM predicts the chemical enrichment by explicitly
    tracking both the histories of star formation and mass-loading. After
    exploring the \textsc{EAGLE} simulation, we discover a compact yet
    flexible model that accurately describes the average star formation
    histories of galaxies. Applying a novel method of chemically measuring
    $\eta$ to \textsc{EAGLE}, we find $\eta$ can be parametrised by its
    dependence on stellar mass and specific SFR as $\log\eta\propto
    M_*^{\alpha}s{\mathrm{SFR}}^{\beta}$, with $\alpha{=}{-}0.12$ and
    $\beta{=}0.32$ in \textsc{EAGLE}. Our chemically-inferred $\eta$ agrees
    remarkably well with the kinematic measurements by Mitchell et al.
    After extensive tests with \textsc{EAGLE}, we apply an NE-CEM Bayesian
    analysis to the SDSS data, yielding a tight constraint of
    $\log(\eta/0.631) =
    0.731{\pm}0.002\times(M_*/10^{9.5}M_{\odot})^{-0.222\pm0.004}
    (s{\mathrm{SFR}}/10^{-9.5}yr^{-1})^{0.078\pm0.003}$, in good agreement
    with the down-the-barrel measurements. Our best-fitting NE-CEM not only
    accurately describes the metallicity-stellar mass-SFR relation at
    $z{=}0$, but also successfully reproduce the so-called ``fundamental
    metallicity relation'' at higher redshifts. Our results reveal that
    different galaxies form stars and enrich their gas in a non-equilibrium
    but strikingly coherent fashion across cosmic time.
\end{abstract}




\vspace{1in}

\section{Introduction}
\label{sec:intro}

The metallicity of the interstellar medium~(ISM) provides a key diagnostic
of the ejective feedback mechanisms, i.e., galactic
outflows~\citep{Heckman1990, Veilleux2005}, in theories of galaxy
formation~\citep{Somerville2015, Naab2017}. In particular, the chemical
enrichment history~(CEH) of a galaxy is shaped by the interplay between
metal production by stellar nucleosynthesis along the star formation
history, metal dilution in the ISM by the accretion of metal-poor
gas, and metal loss due to the ejection of metal-enriched material by
outflows~\citep{Larson1972, Tinsley1980, Dekel1986, Low1999, Dalcanton2007,
Finlator2008, Andrews2017, Weinberg2017b}. In this paper, by examining such
complex interplay in the \eagle{} hydrodynamical
simulation~\citep{Schaye2015, Crain2015}, we develop a simple yet
comprehensive chemical evolution model~(CEM) to simultaneously reconstruct
the average SFH and constrain the physics of galactic winds driven by
stellar feedbacks, from the present-day metallicity~($\zgas$), stellar
mass~($\ms$), and star formation rate~(SFR; $\sfr$) of galaxies observed by
the Sloan Digital Sky Survey~\citep[SDSS;][]{York2000}.

One of the primary goals of galaxy CEMs in the literature is to provide an
analytic framework for interpreting the observed scaling relations between
the gaseous metallicity $\zgas$~(as measured by log $\mathrm{O/H}$, the
oxygen-to-hydrogen abundance ratio in the ISM) and other physical
properties of galaxies, as well as the redshift evolution~(or lack thereof)
of these relations~\citep[][and references therein]{Maiolino2019}. For
instance, the positive correlation between oxygen abundance and stellar
mass of galaxies~(a.k.a., the mass-metallicity relation; MZR) has been
observed in both the local Universe~\citep{Lequeux1979, Tremonti2004,
Zahid2011, Andrews2013, Gao2018, Huang2019} and at higher redshifts of
$z{\sim}1{-}3$~\citep{Savaglio2005, Erb2006, Maiolino2008, Mannucci2009,
Henry2013, Sanders2021, Wang2022, Li2022}. The MZR is likely driven by the
anti-correlation between $\ms$ and the mass-loading of outflows $\eta$,
defined as
\begin{equation}
    \eta = \frac{\mathrm{mass\;outflow \;rate}}{\mathrm{star \;formation
    \;rate}} = \frac{\mout}{\sfr},
\end{equation}
because the outflowing material is less likely to escape the gravitational
potential a massive system than a dwarf galaxy~\citep{Peeples2011}.
Furthermore, the scatter in the MZR~($\sim$0.1 dex) may be driven by a
third parameter, including gas fraction~\citep{Hughes2013, Bothwell2016,
Brown2018, Zu2020, Chen2022}, size~\citep{Ellison2008, Yabe2014}, and
SFR~\citep{Mannucci2010, Lopez2010, Yates2012, Andrews2013}.  In
particular, SDSS galaxies form a tight~($\sim$0.05 dex) surface $\zgas(\ms,
\sfr)$ in the 3D parameter space of $\zgas$, $\ms$, and $\sfr$, dubbed the
``fundamental metallicity relation''~\citep[FMR;][]{Mannucci2010,
Lopez2010}. Intriguingly, the observed FMR exhibit little evolution
from $z=0.1$ up to $z{=}2.5$~\citep{Mannucci2010, Topping2021}, despite a
significant redshift evolution in the amplitude of
MZR~\citep[$\log\mathrm{O/H}\propto (1+z)^{-2.3}$ at fixed mass;][]{Ly2016}.

Although the existence of FMR remains a subject of intense
debate~\citep{Sanchez2013, BarreraBallestero2017, Cresci2019}, various
theoretical models have subsequently been proposed to explain the
phenomenon and its apparent lack of redshift evolution~\citep{ Dayal2013,
Forbes2014, Harwit2015, Hunt2016, Kacprzak2016}.  The most important among
them is a class of CEMs named the ``equilibrium''
 or ``gas-regulator'' models~\citep{Bouche2010, Dave2012, Lilly2013}.
The equilibrium CEMs assume a constant or slowly-evolving gas reservoir,
which implies that the gas accretion rate maintains an instantaneous
balance with the rate of gas consumption due to star formation and
mass-loaded outflows.  In essence, the equilibrium CEMs drive the galaxies
to chemical equilibrium so rapidly that the observed metallicity in the ISM
has little memory of the past histories of star formation or chemical
enrichment, establishing a tight $\zgas(\ms, \sfr | z)$ relation at each
epoch.  Alternatively, however, the apparent constancy of FMR with redshift
may emerge out of a non-equilibrium yet somewhat coherent enrichment of
star-forming galaxies on the $\ms$ vs. $\sfr$ plane.  To explore a more
general CEM that does not assume a steady-state gas reservoir, we develop a
non-equilibrium CEM~(NE-CEM) that explicitly tracks the variation of metal
production, dilution, and ejection over the history of star-forming
galaxies.

The success of such an NE-CEM depends critically on the accuracy of {\it
in-situ} SFHs reconstructed from observations.  For distant galaxies with
unresolved stellar populations, spectral energy distribution~(SED) fitting
remains the only viable means of reconstructing SFHs~\citep{Kauffmann2003,
Walcher2011, Conroy2013, Leja2019}, but the large stochasticity in star
formation events renders such reconstruction extremely challenging for
individual galaxies~\citep{Broussard2019, Tacchella2020, Iyer2022,
WangYang2022}. The issue can be potentially circumvented by reconstructing
the {\it average}~(hence smooth) SFH for a large number of {\it similar}
galaxies, and the lack of sudden bursts is not important for modelling pure
core-collapsed supernova~(CCSN) elements~\citep[e.g.,
oxygen;][]{Woosley1995, Johnson2019} due to the short lifetimes of CCSN
progenitors~\citep[see][for the impact of bursts on various abundance
ratios]{Johnson2020}. Conceptually speaking, the average SFH is the
convolution between the average (baryon) mass accretion history~(MAH) of
their host haloes and a transfer function that characterises the delayed
star formation episodes on shorter timescales~\citep{Wang2019}.  By
studying the variability of SFHs in various hydrodynamical simulations and
semi-analytic models~(SAMs), \cite{Iyer2020} found that the {\it in-situ}
SFHs in different galaxy formation models are all coherent with the MAHs of
their parent haloes on long timescales~(${>}3$ G$yr$).  Therefore, given
that galaxies with the same $\ms$ and $\sfr$ observed at $z{=}0$ should
live in similar haloes~\citep{Zu2015, Zu2016, Zu2018}, and that halo MAHs
follow a simple, universal profile~\citep{vandenBosch2002, Zhao2009}, we
expect the {\it average} {\it in-situ} SFH of those galaxies to be fairly
representative of the individual ones.  In the current work, we aim to show
that using a well-motivated functional form predicted by the \eagle{}
simulation allows us to observationally reconstruct robust average {\it
in-situ} SFHs for our NE-CEM analysis.

The key to constraining the physics of stellar feedback-driven outflows
lies in robustly measuring the dependence of $\eta$ on galaxy properties
along the SFH, i.e., the mass-loading history~(MLH) of galactic outflows.
Although there exists a plethora of observations that either directly
caught outflows in action from the disks of star-forming
galaxies~\citep[][and references therein]{Rupke2018}, or provided indirect
evidence via the detection of a significant amount of both metals in the
circumgalactic medium~\citep{Peeples2014, Werk2014, Tumlinson2017} and dust
in the intergalactic medium~\citep{Menard2010, Zu2011} around active and
even quiescent galaxies~\citep{Zhu2014, Huang2016, Zu2021}, it is
challenging to accurately measure the mass outflow rate $\mout$ due to the
various uncertainties associated with converting column density and wind
velocity of some outflowing component~(i.e., hot, warm, or cold) into a
total mass outflow rate~\citep{Murray2007, Chisholm2016}. Furthermore,
direct measurements of $\eta$ usually require ``down-the-barrel''
observations of rest-frame ultraviolet~(UV) absorption lines from
space~\citep{Heckman2015, Chisholm2018} or deep narrowband imagining of
H$\alpha$ emission~\citep{Mcquinn2019}, which are mostly limited to
low-redshift systems. In this paper, we adopt an indirect and complementary
approach and reconstruct the MLHs of galactic outflows over the entire
lifetime of star-forming galaxies from their observed $\zgas(\ms, \sfr)$
relation at $z{=}0$. Regardless of the epoch, the mass-loading factor of a
galaxy should depend on its stellar mass $\ms$, which sets the depth of
gravitational potential, and the specific star formation rate $\ssfr$,
which controls the valve of energy and momentum-injection due to stellar
feedback. Therefore, we parametrise $\eta$ at any given epoch as a function
of $\ms$ and $\ssfr$ of the galaxy at that epoch, which we are able to
predict from the reconstructed SFH in our NE-CEM.

This paper is organised as follows.  We develop and calibrate our analytic
models of the SFH and MLH by investigating the \eagle{} simulation in
\S\ref{sec:ceh}. We then build our NE-CEM by combining the SFH and MLH
models and demonstrate the efficacy of NE-CEM using the mock data from
the \eagle{} simulation in \S\ref{sec:necem}.  By performing a first-cut
NE-CEM analysis using the SDSS metallicity-stellar mass-SFR relation, we
derive stringent constraints on the dependence of mass-loading factor on
$\ms$ and $\ssfr$ in \S\ref{sec:sdss}.  We also discuss the physical
implication of our constraints and the physical cause of the so-called
fundamental metallicity relation, before concluding our paper and looking
to the future in \S\ref{sec:conc}.  Throughout this paper, we assume a flat
Universe with $\Omega_m{=}0.306$ and $h{=}{0.677}$ for distance and age
calculations. We indicate the base-10 logarithm with $\lg$, and use $\sfr$
and $\mathrm{SFR}$ interchangeably to refer to the star formation rate in
the main text and figures, respectively.

\section{Chemical enrichment history of galaxies in the \eagle{} simulation}
\label{sec:ceh}

\begin{figure*}
\centering\includegraphics[width=0.96\textwidth]{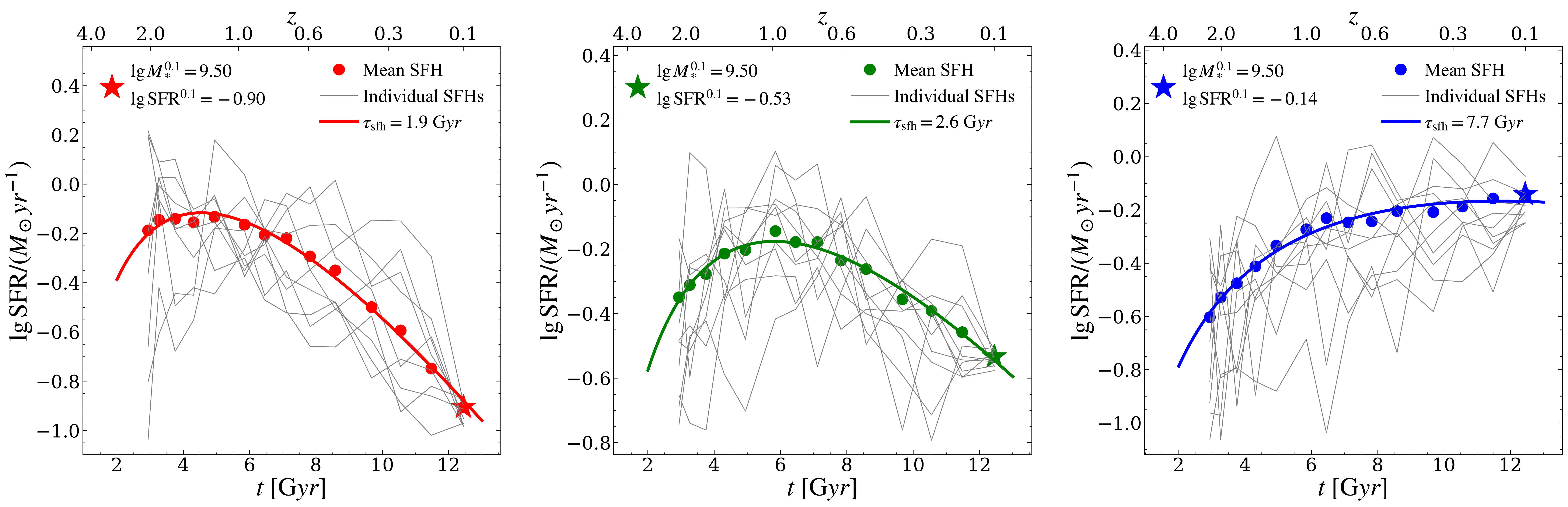}
    \caption{Star formation histories of galaxies in the \eagle{}
    simulation. The three panels are for galaxies observed with the same
    $\ms{=}10^{9.5}\msol$ but different $\sfr$ at $z{=}0.1$~(i.e.,
    $\mathrm{SFR}^{0.1}$), indicated by the star symbols at
    $t{=}12.45\,\gyr$.  In each panel, the top and bottom x-axes indicate
    the redshift and time since the big bang, respectively.  Thin gray
    lines shows the individual SFHs of ten random galaxies within the same
    2D bin of $\ms^{0.1}$ and $\mathrm{SFR}^{0.1}$, while circles represent
    the mean SFH of all galaxies within that bin. The mean SFH can be well
    described by the $\texttt{powexp}$ model~(solid curve with colour),
    with the best-fitting value of the SFH timescale $\tausfh$ listed in
    the top right corner.  \label{fig:sfheagle}}
\end{figure*}

In order to build a robust CEM for galaxies that are not necessarily in
equilibrium, we start by systematically investigating the CEH of
star-forming galaxies in the \eagle{} hydrodynamical simulation.  In
particular, we firstly develop a compact yet flexible model for the average
{\it in-situ} SFHs of galaxies measured from the \eagle{} simulation in
\S\ref{subsec:sfh}, and then solve the MLHs of galactic outflows by
applying a standard one-zone CEM with inflows and outflows~(i.e., open-box)
along galaxy SFHs in \S\ref{subsec:cem}.

The \eagle{} suite of cosmological simulations~\citep{Crain2015,
Schaye2015} constitutes a set of hydrodynamical simulations run with
different box sizes, particle numbers, and sub-grid physics. In this work
we employ the ``Ref-L100N1504'' simulation, which has a periodic box size
of 100 $\mpc$~(comoving) and a particle number of $2{\times}1504^3$~(i.e.,
equal number of dark matter and baryonic particles). As a ``reference''
model, this particular simulation implemented a sub-grid feedback
prescription that was calibrated to reproduce the galaxy stellar mass
function observed by SDSS at $z{=}0.1$~\citep{Schaye2015, Furlong2015}.
For further details of the sub-grid implementation in \eagle{}, we refer
interested readers to \citet{Schaye2008} for star formation,
\citet{DallaVecchia2012} for stellar feedback and galactic outflows, and
\citet{Wiersma2009} for metal enrichment, respectively.

We are primarily concerned with the CEH modelling for star-forming galaxies
in the low-to-intermediate stellar mass range~($\ms{<}10^{10.2}\msol$),
where galactic outflows are driven by stellar feedbacks.  The star-forming
galaxies in the Ref-L100N1504 simulation are broadly consistent with
observations at $\ms{<}\mathrm{a\,few}\times 10^{10}\,\msol$.  In
particular, the predicted fraction of star-forming galaxies in the
simulation agrees well with the observations at $z{=}0.1$; The predicted
$\ssfr$ of those galaxies are lower than the observations by
$0.2{-}0.3$ dex but within the expected systematic errors due to, e.g., the
unknown initial mass function~(IMF); The metallicity scaling relations
predicted by the \eagle{} sub-grid model is qualitatively consistent with
observations, though the exact slope of the MZR depends on
resolution~\citep{DeRossi2017}.  The overall agreement becomes slightly
worse for star-forming galaxies at the higher mass where Active Galactic
Nuclei~(AGNs) start to dominate the feedbacks. Therefore, the
Ref-L100N1504 simulation provides an excellent laboratory for calibrating
and testing the SFH and MLH of our analytic NE-CEM regulated by stellar
feedbacks. We will refer to the ``Ref-L100N1504'' simulation simply as the
``\eagle{}'' simulation for the rest of the paper.

\subsection{Star formation histories of \eagle{} galaxies}
\label{subsec:sfh}

As emphasized in the Introduction, a robust reconstruction of the average
{\it in-situ} SFH is key to building an accurate NE-CEM that tracks the
production, dilution, and ejection of metals in the ISM.  In particular,
metal production is directly set by the {\it in-situ} SFH multiplied by the
stellar yield, dilution is associated with the gas content that is tied
with the {\it in-situ} SFH via the empirical star-forming
law~\citep{Kennicutt1998}, and ejection is in sync with the {\it in-situ}
SFH via the energy and/or momentum injected after each star formation
episode. Therefore, we will measure the individual SFHs for the
$z{=}0.1$~($t_{0.1}{=}12.45\gyr$) galaxies in the \eagle{} simulation, in
hopes of finding a robust model for the average SFHs. For the sake of
brevity, from now on we will refer to the ``{\it in-situ} SFH'' simply as
``SFH''.

\subsubsection{Extracting individual SFH from the merger tree}
\label{subsubsec:mtree}

In order to measure the individual SFHs, we first build a merger tree for
each galaxy in the $z{=}0.1$ output~(i.e., snapshot 27) by tracking all of
its progenitors in the previous 26 snapshots. In the \eagle{} simulation,
each halo~(including both main and sub-haloes) in \eagle{} has a unique
\texttt{GalaxyID} and a \texttt{DescendantID} that points to the
\texttt{GalaxyID} of its direct descendant in the next snapshot. In an
event of a merger, multiple haloes would share the same
\texttt{DescendantID}. Since we do not need to explicitly track merger
activities in our NE-CEM, we combine multiple progenitors at the same epoch
as one synthetic star-forming conglomerate along the SFH.  That is, we sum
the amount of {\it in-situ} star formation and stellar mass in all its
progenitors at the same epoch to obtain the values of $\sfr$ and $\ms$,
respectively, for that epoch. In other words, we ``collapse'' the multiple
sub-branches of the merger tree into the main branch, and then measure a
total SFH from that single branch. Since major mergers experienced by the
star-forming galaxies are rare, the main branch dominates the
merger tree in the \eagle{} simulation, on average accounting for 96\% and
74\% of the total stellar mass at $z{=}0.5$ and $z{=}2.0$, respectively.

\begin{figure*}
\centering\includegraphics[width=0.96\textwidth]{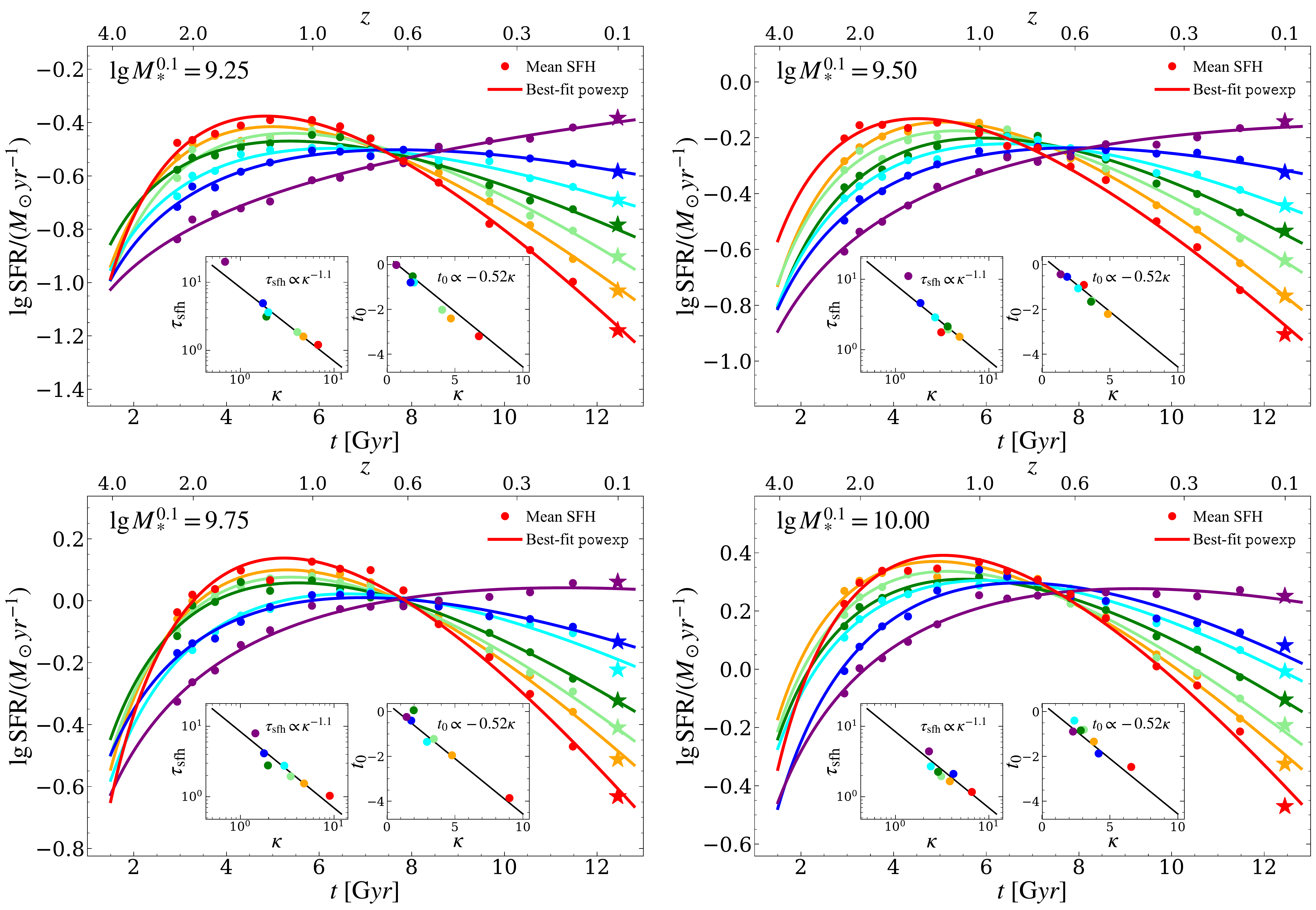}
\caption{The mean SFHs of \eagle{} galaxies with $\lg\ms{=}9.25$~(top
    left panel), $9.5$~(top right), $9.75$~(bottom left) and $10$~(bottom
    right) observed at $z{=}0.1$.  In each panel, the mean SFHs of galaxies
    with seven different SFRs observed at $z{=}0.1$~(star symbols at
    $t{=}12.45\gyr$) are indicated by the circles with seven different
    colours~(increasing $\mathrm{SFR}^{0.1}$ from red to purple), while
    curves of the matching colours show the best-fitting $\texttt{powexp}$
    models of SFH. The top and bottom x-axes indicate the redshift and age
    of the Universe, respectively.  The left and right inset panels show
    the dependencies of the best-fitting $\tausfh$ and $t_0$, respectively,
    on $\kappa$ for each of the seven SFHs shown in the main panel~(circles
    with matching colours). Solid line in the left~(right) inset panel
    shows the best-fitting power-law~(linear) fit, which is the same across
    all the stellar mass bins.  \label{fig:sfheagleall}}
\end{figure*}

Figure~\ref{fig:sfheagle} shows the SFHs measured from the \eagle{}
simulation for galaxies with three different $\sfr$ but the same
$\lg\ms{=}9.5$ observed at $z{=}0.1$~(marked by the star symbols). In each
panel, the SFHs of ten random galaxies of the same ($\ms^{0.1}$,
$\sfr^{0.1}$) are shown as individual thin gray lines~(with the superscript
$0.1$ indicating quantities observed at $z{=}0.1$), while the circles are
the mean SFH of these galaxies. As expected in the Introduction, despite
the large stochasticity exhibited by individual SFHs, the average SFH of
galaxies with the same ($\ms^{0.1}$, $\sfr^{0.1}$) appears smooth and can
thus be well described by an analytic function, indicated by the thick
colored curve in Figure~\ref{fig:sfheagle}~(as will be discussed further
below).

\begin{figure*}
\centering\includegraphics[width=0.8\textwidth]{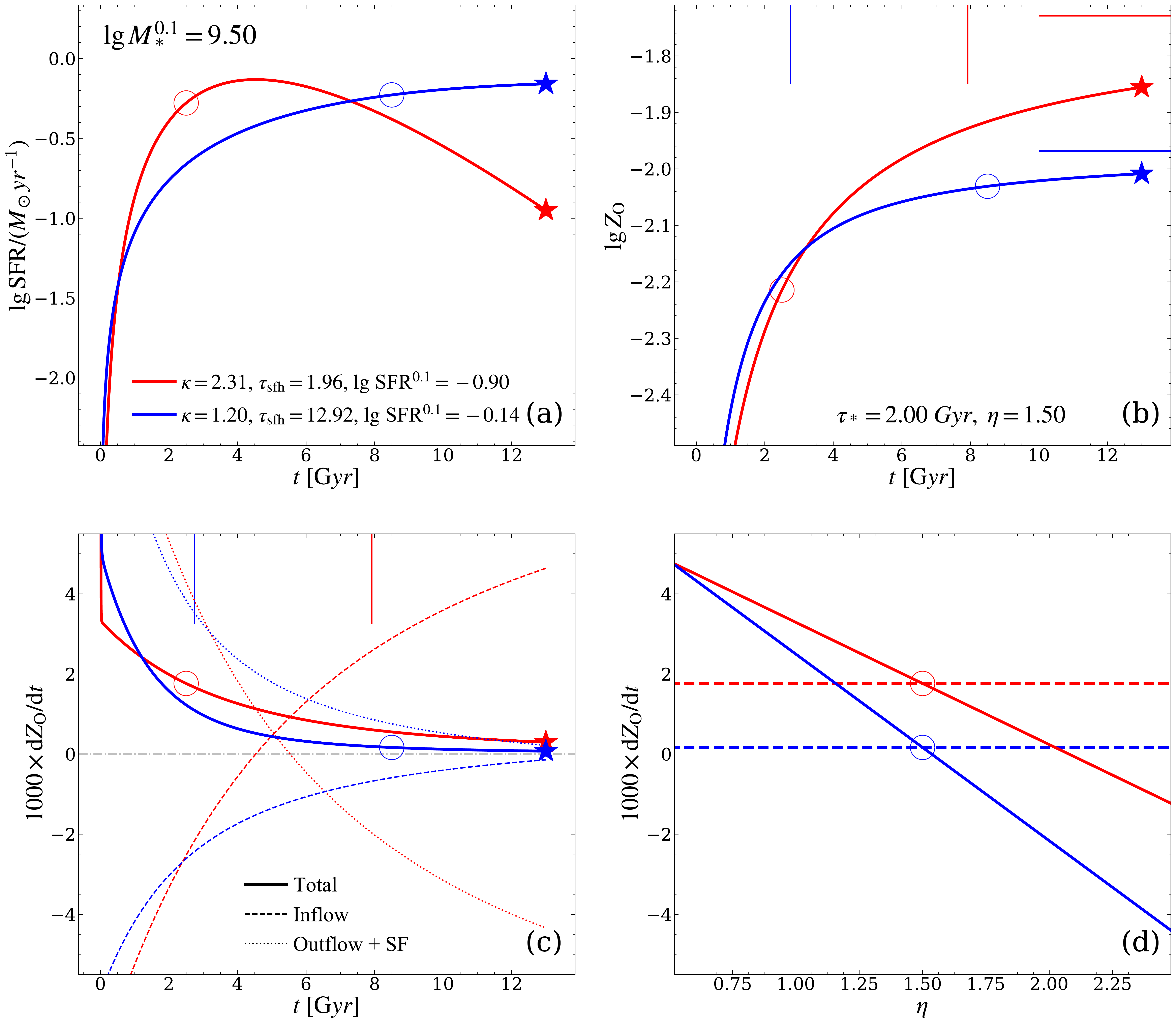}
    \caption{A pedagogical illustration of the different histories of star
	formation~(top left), chemical enrichment~(top right), and rate of
    enrichment $\dzodt$~(bottom left) between a typical ``young''
	galaxy~(blue) and an ``old'' galaxy~(red) with the same stellar
	mass of $10^{9.5}\msol$ observed at $z{=}0.1$, assuming
    the \powexp{} SFH model with constant
	gas consumption timescales and mass-loading factors. Bottom right
	panel illustrates the methodology of our
    mass-loading reconstruction using two arbitrary epochs for the
    young~(blue circle at $t{=}8.5\,\gyr$) and old~(red circle at
    $t{=}2.5\,\gyr$) galaxies.  {\it Panel (a)}: Two example \powexp{} SFHs
    with the parameters listed in the bottom, normalised so that the two
    galaxies arrive at the same stellar mass of $10^{9.5}\msol$ by
    $t{=}12.45\,\gyr$ but with different SFRs~(stars).{\it Panel (b)}: The
    two evolution of oxygen abundance $\zo$ as functions of time,
    predicted by Equation~\ref{eqn:zosimple} with $\tau_*{=}2\,\gyr$ and
    $\eta{=}1.5$.  Horizontal and vertical short lines indicate the
    equilibrium metallicities~($\lg\zoeq{=}{-1.97}$ vs. ${-1.73}$ for red
    vs. blue) and timescales~($2.75$ vs. $7.92\,\gyr$) predicted by
    Equation~\ref{eqn:zoeq} and~\ref{eqn:ttilde}, respectively.  {\it Panel
    (c)}: The total change rates~(thick solid curves) of $\zo$ as functions
    of time predicted by Equation~\ref{eqn:dzodtpowexp}, each of which can
    be decomposed into an ``Inflow'' component~(dashed) and an
    ``Outflow+SF'' component~(dotted). See text for detail.
     {\it Panel (d)}: Solid lines are the change rates of oxygen abundance
     predicted by Equation~\ref{eqn:dzodtpowexp} for different values of
     $\eta$. Horizontal dashed lines indicate the actual values of $\dzodt$
     at the two epochs indicated by the circles in panels (a), (b), and
     (c). The intersection points between the solid and dashed lines
     indicate the correct values of $\eta{=}1.5$.
    \label{fig:ceh_simple} }
\end{figure*}

More important, the average SFHs of galaxies with different $\sfr^{0.1}$
exhibit distinct shapes from one another despite arriving at the same $\ms$
at $z{=}0.1$. In particular, galaxies with higher $\sfr^{0.1}$ tend to have
a later peak of star formation~(compare left and middle panels of
Figure~\ref{fig:sfheagle}), while those with the highest $\sfr^{0.1}$ have
yet to reach the peak~(right panel of Figure~\ref{fig:sfheagle}).  This
coherent shape variation of the average SFH with $\sfr^{0.1}$ indicates
that the individual SFH is non-Markovian --- the {\it instantaneous} star
formation at the observed epoch {\it does} retains some memory of its
star-forming history in the past. This non-Markovian property of SFH is
likely associated with the coherence of SFH with the MAH of haloes on long
timescales, a generic prediction of structure formation in the $\Lambda$
Cold Dark Matter~($\lcdm$) cosmology. Therefore, we expect that such
non-Markovian property of SFH is present not only in the \eagle{}
simulation but also in the real Universe, making it plausible to robustly
reconstruct the average SFH for SDSS galaxies observed with the same $\ms$ and
$\sfr$.

\subsubsection{Modelling average SFH of galaxies at fixed ($\ms$, $\sfr$)}
\label{subsubsec:avgsfh}

The reconstruction of SFH from observations requires an accurate fitting
formula for the average SFH. After extensive tests, we find that the
``powerlaw-exponential''~(hereafter shortened as ``\powexp'') functional form
of Equation~\ref{eqn:sfh}, which resembles the Schechter function widely
used for describing galaxy luminosity functions~\citep[see
also][]{Katsianis2021}, provides excellent description of the average SFHs
measured from \eagle{}. It consists of an early power-law increase and a
late exponential decline, so that
\begin{equation}
    \sfr(t) = \dot{M}_{*,0}\left(\frac{t-t_0}{\tausfh}\right)^{\kappa}
\exp \left(\frac{t_0-t}{\tausfh}\right)
    \label{eqn:sfh}
\end{equation}
where $\dot{M}_{*,0}$ determines the overall amplitude of star formation,
$t_0$ is the starting time of star formation, $\kappa$ sets the slope of
the rapid increase at the onset of star formation, and $\tausfh$ is the
characteristic timescale of the exponential decline at late times.  Thick
solid curves in Figure~\ref{fig:sfheagle} show the best-fitting models of
Equation~\ref{eqn:sfh}, with the best-fitting values of $\tausfh$ indicated
on the top right of each panel. As expected, the characteristic timescale
$\tausfh$ increases monotonically with $\sfr^{0.1}$, yielding
$\tausfh{=}1.9\,\gyr$, $2.6\,\gyr$, and $7.7\,\gyr$ for
$\lg\sfr^{0.1}{=}-0.9$, $-0.53$, and $-0.14$, respectively.

Figure~\ref{fig:sfheagleall} demonstrates the efficacy of our \powexp{} SFH
model in describing the average SFHs of \eagle{} galaxies with
$\lg\ms^{0.1}{=}9.25$, $9.5$, $9.75$, and $10$ in the four panels~(arranged
by increasing $\ms$ from the top left to the bottom right), with each panel
showing the SFHs for galaxies with seven different $\sfr^{0.1}$ at that
$\ms$~(increasing $\sfr^{0.1}$ from red to purple, as indicated by the
stars at $z{=}0.1$). Similar to Figure~\ref{fig:sfheagle}, solid circles
are the average SFHs measured from \eagle{}, while solid curves of the same
colour indicate the best-fitting \powexp{} model of Equation~\ref{eqn:sfh}.
Overall, the best-fitting \powexp{} models provide excellent description of
the average SFHs measured directly from the \eagle{} simulation for
star-forming galaxies with $\ms^{0.1}$ between roughly $10^9\msol$ to
$10^{10}\msol$, similar to the stellar mass range that we aim to explore in
the SDSS data. We do not extend our model to galaxies of even higher $\ms$
because we want to limit our analysis to the regime of stellar feedbacks,
while the outflows in those high-$\ms$ systems are progressively driven by
AGNs.

The \powexp{} SFH model of Equation~\ref{eqn:sfh} has four free parameters,
while in the observation we usually have only two measured quantities,
i.e., $\ms$ and $\sfr$ at the observed epoch. Ideally, one would stack the
spectra of the observed galaxies at fixed $\ms$ and $\sfr$, and apply the
\powexp{} SFH to SED-fitting techniques to derive the other two parameters
$\kappa$ and $\tausfh$~(Chen et al. {\it in prep}).  However, for the
purpose of our first-cut analysis, it would be useful to find an empirical
constraint to reduce the number of degrees of freedom without resorting to
sophisticated machineries like SED-fitting.

The inset panels in Figure~\ref{fig:sfheagleall} point to a promising path
to such an empirical constraint. In each panel, the left and right inset
panels show the relations between the best-fitting values of $\kappa$ and
those of $\tausfh$ and $t_0$, respectively. Note that we allow the values
of $t_0$ to be negative to better fit the shapes of the SFHs at $z{\sim}2$,
above which we cannot obtain meaningful $\sfr$ measurements from the
simulation. The colours of the circles are matched to those of the
best-fitting SFH curves in the main panel. Black solid lines are the same
across all the inset panels of $\tausfh$ vs. $\kappa$, indicating the
best-fitting power-law relation $\tausfh{=}{0.92} \kappa^{-1.1}$.
Likewise, black solid lines in the inset panels of $t_0$ vs. $\kappa$ are
the best-fitting line relation $t_0 {=} {-}0.52 \kappa {+} 0.37$.  Since
both black solid lines provide reasonably good fits to the respective
relations between best-fitting parameters, we assume that the three
parameters of Equation~\ref{eqn:sfh} roughly follow a power-law relation
\begin{equation}
    \lg\tausfh = A_{\tau} \lg \kappa + B_{\tau},
    \label{eqn:taukappa}
\end{equation}
and a linear relation
\begin{equation}
    t_0 = A_0 \kappa + B_0,
    \label{eqn:t0kappa}
\end{equation}
simultaneously. In essence, galaxies that start forming stars earlier tend
to experience faster growth during the power-law
phase~(Equation~\ref{eqn:t0kappa}), and then more rapid declines in the
exponential phase~(Equation~\ref{eqn:taukappa}). Such tendency of galaxy
SFHs is consistent with that of halo MAHs in $\lcdm$, where older haloes
usually experience faster early-time growth when the Universe was dense
than their younger counterparts of the same mass~\citep{Zhao2009}.
Therefore, we expect the power-law~(Equation~\ref{eqn:taukappa}) and linear
scaling relations~(Equation~\ref{eqn:t0kappa}) to be roughly applicable in
the real Universe, but likely with a different set of $\{A_{\tau},
B_{\tau}, A_0, B_0\}$.

\begin{figure}
    \hspace{-0.5cm}\includegraphics[width=0.48\textwidth]{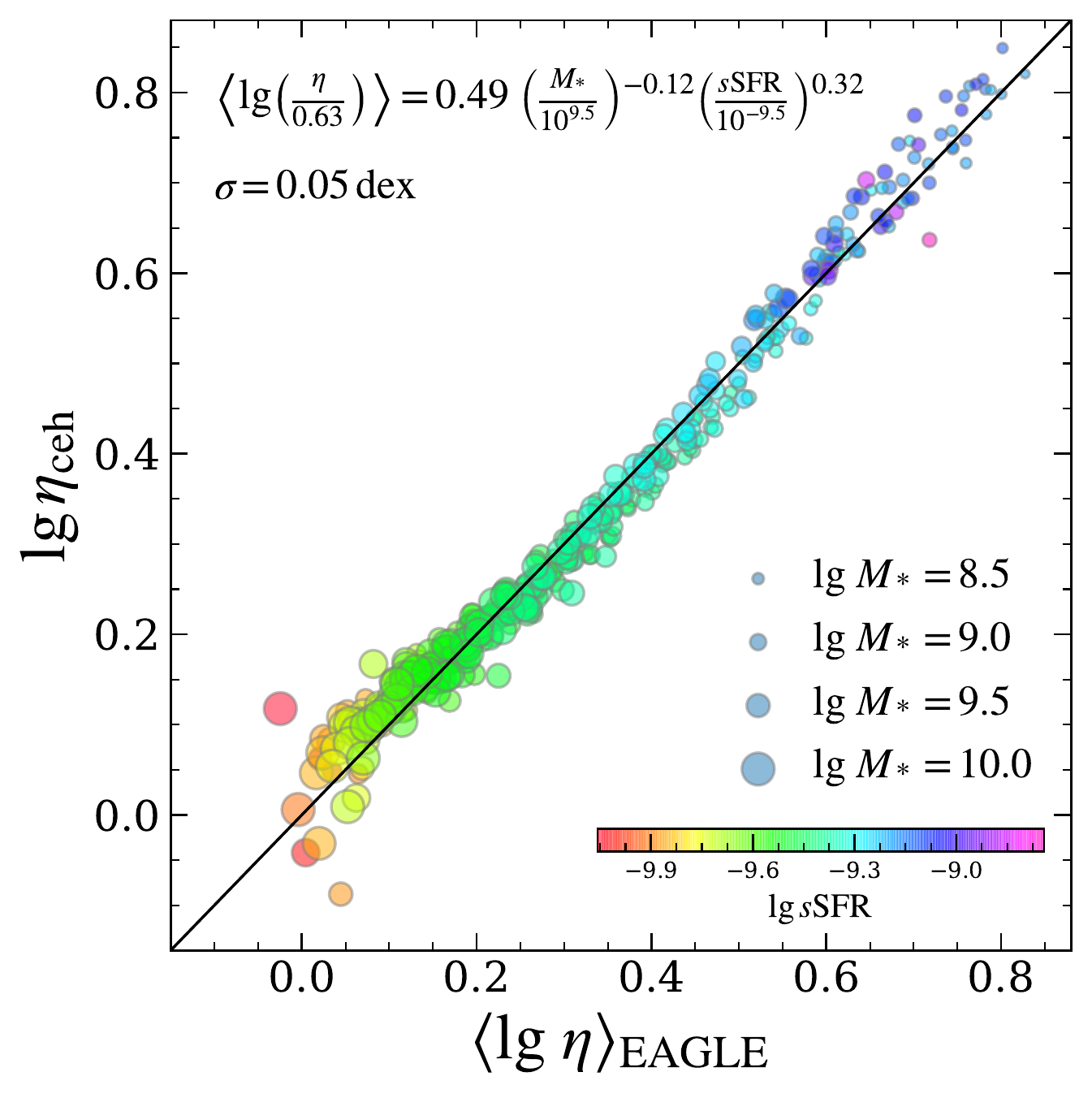}
    \caption{Comparison between the average mass-loading factors
    reconstructed from CEHs~(y-axis) and that predicted from the
    best-fitting model~(x-axis) of
    Equation~\ref{eqn:etacehmodel}~(indicated by the equation in the top
    left corner) for the \eagle{} galaxies. The colour and size of each
    circle represent the $s\mathrm{SFR}$ and $\ms$ of each galaxy, indicated
    by the colourbar and legends, respectively, in the bottom right. The
    circles are consistent with the solid diagonal line that indicates
    the one-to-one relation with a scatter of $0.05$ dex.
    \label{fig:etafit}}
\end{figure}

Finally, given that the average $\ms$ and $\sfr$ of a large sample of
galaxies at some observed epoch $\tobs$ are usually well measured~(e.g., as
$\msobs$ and $\sfrobs$), we can change the variables so that
\begin{equation}
    \sfr(t) = \sfrobs
    \left(\frac{t-t_0(\kappa))}{\tobs - t_0(\kappa)}\right)^\kappa
    \exp \left(\frac{\tobs- t}{\tausfh(\kappa)}\right),
    \label{eqn:sfhalt}
\end{equation}
where we make use of Equation~\ref{eqn:taukappa} and \ref{eqn:t0kappa} to
obtain $\tausfh(\kappa)$ and $t_0(\kappa)$, respectively. On the other
hand, $\kappa$ can be separately derived from the $\ssfr$ at $\tobs$.
Analytically integrating Equation~\ref{eqn:sfh}, we have
\begin{equation}
    \frac{\msobs}{\sfrobs}=\frac{(1-r)\tausfh^{\kappa+1}}{(\tobs-t_0)^{\kappa}}
    \frac{\gamma\left(\kappa+1, \frac{\tobs-t_0}{\tausfh}\right)}
    {\exp \left(-\frac{\tobs- t_0}{\tausfh}\right)}
    \equiv \mathcal{F}(\kappa),
    \label{eqn:kappamap}
\end{equation}
where $\gamma$ is the incomplete Gamma function and $r$ is the IMF-averaged
recycle fraction, defined as the fraction of mass formed into stars that is
returned to the ISM by supernovae and evolved stars. The value of $\kappa$
can thus be solved trivially from the inverse function of $\mathcal{F}$ as
$\kappa = \mathcal{F}^{-1}(\msobs/\sfrobs)$. In this way, we can
analytically derive the average SFH from any combination of $\msobs$ and
$\sfrobs$, for any given set of $\{A_\tau, B_\tau, A_0, B_0\}$.

\subsection{Chemical Evolution along the average SFH}
\label{subsec:cem}

\subsubsection{Standard CEM with the \powexp{} SFH}
\label{subsubsec:cempowexp}

Armed with the \powexp{} SFH model developed in~\S\ref{subsec:sfh}, we are
now able to analytically track the oxygen abundance in the ISM by applying
an open-box CEM along any given \powexp{} SFH. In the standard
instantaneous recycling approximation and the notations of
\citet{Weinberg2017b}, the evolution equation for the total mass of oxygen
in the ISM $\mo$ is
\begin{equation}
    \modot= \mocc \sfr - (1-r)\zo \sfr -\eta \zo\sfr,
    \label{eqn:modot}
\end{equation}
where $r$ is the recycle fraction, $\zo{\equiv}\mo/\mgas$ is the current
oxygen abundance by mass in the ISM, $\mocc$ is the IMF-averaged oxygen
yield, defined as the mass of oxygen produced and returned to the ISM per
solar mass of star formation~(i.e., oxygen yield per stellar generation, as
opposed to the net yield $y_\mathrm{O}$). For the Chabrier stellar IMF
assumed in \eagle{}, we adopt recycle fraction $r{=}0.4$ and oxygen yield
$\mocc{=}0.022$\footnote{We adopt an upper mass cutoff of the Chabrier IMF
as $m_{\mathrm{up}}{=}80\,\msol$, and derive a net yield of
$y_\mathrm{O}{=}0.0367$ based on the Fig. 5 of \citet{Vincenzo2016}. We
then multiply $y_\mathrm{O}$ by $(1-r){=}0.6$ to obtain $\mocc{=}0.022$. We
note that as \citet{Griffith2021} pointed out, the value of $\mocc$ could
vary by a factor of three even at fixed IMF depending on the assumptions
about blackhole formation.}. In Equation~\ref{eqn:modot}, the first term
represents oxygen production rate by core-collapsed supernovae~(CCSNe), the
second term combines the depletion rate of oxygen previously in the ISM
into stars and the recycling rate of oxygen originally locked in the stars
back into the ISM, and the third term describes the ejection of oxygen by
galactic outflows, with a mass-loading factor $\eta$, the key parameter
that we want to constrain in this paper.

\begin{figure*}
\centering\includegraphics[width=0.96\textwidth]{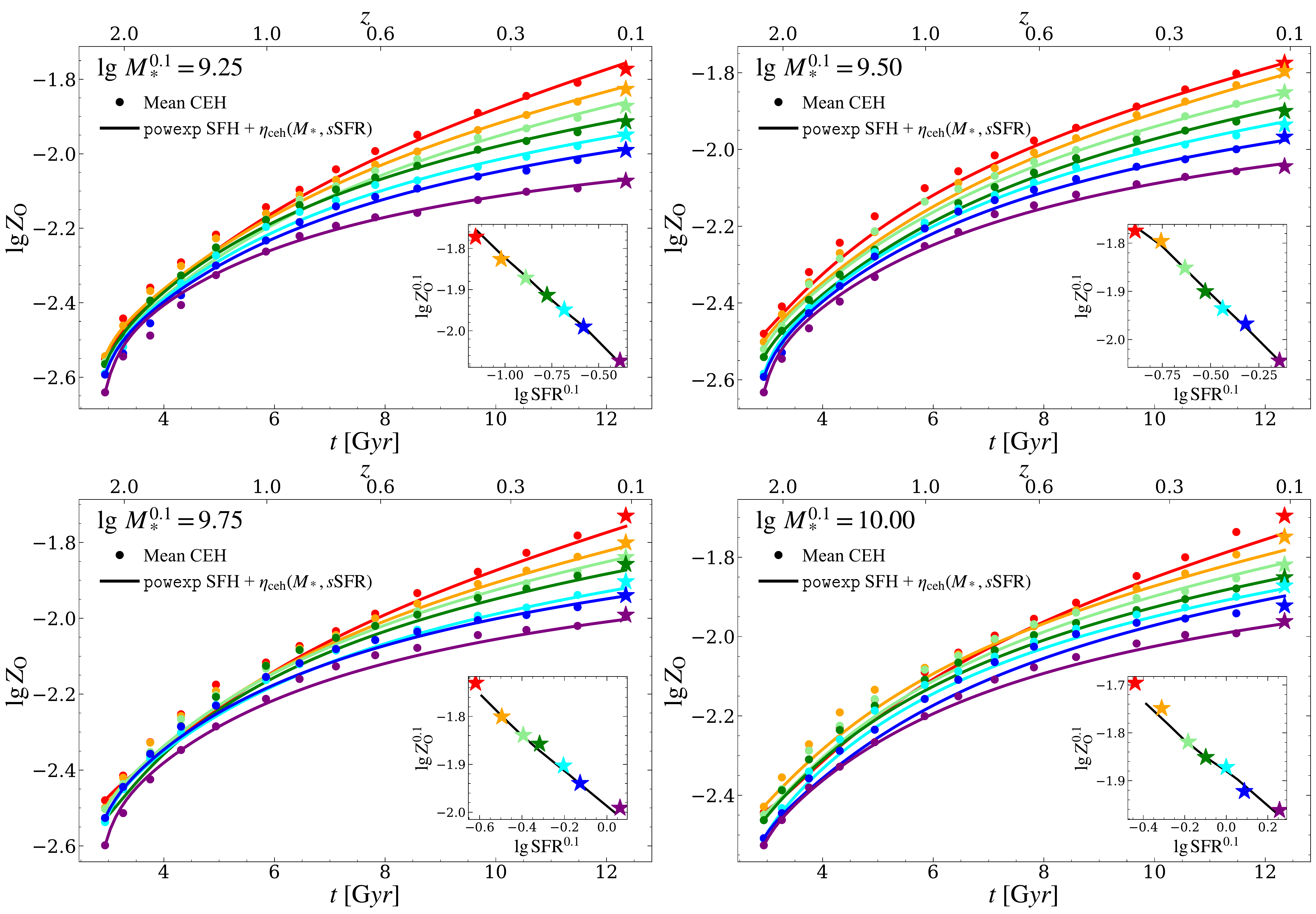}
    \caption{Similar to Figure~\ref{fig:sfheagleall}, but for the CEHs. In
    each panel, coloured circles are the mean CEHs measured from the
	\eagle{} simulation, which can be well described by the solid
	curves with matching colours, predicted by the combination of the
	best-fitting models of \powexp{} SFHs and $\etaceh$ from
	Equation~\ref{eqn:etacehmodel}. Inset panels highlight the
	agreement between the mock measurements from the simulation~(stars) and
	the best-fitting model predictions~(solid black line) at $z{=}0.1$.
\label{fig:ceheagleall} }
\end{figure*}
0
Although Equation~\ref{eqn:modot} assumes the metallicity of the ejecta
$\zo^{\mathrm{ej}}$ is the same as the ISM~(i.e., outflowing gas is pure
ISM), over-enriched outflows~(i.e., ISM entrainment fraction is below
unity) can be accounted for by substituting the metal-loading
factor $\zeta$~\citep{Peeples2011}
\begin{equation}
    \zeta \equiv \frac{\zo^{\mathrm{ej}}}{\zo} \frac{\mout}{\sfr}
    \label{eqn:metalloading}
\end{equation}
for $\eta$ in our formulae.

Considering an evolving gas reservoir, the change rate of oxygen abundance
is thus
\begin{equation}
    \frac{\dd \zo}{\dd t}= \frac{\modot}{\mgas} -
    \frac{\dot{M}_{\mathrm{gas}}}{\mgas}\zo,
    \label{eqn:dzodtsimple}
\end{equation}
where the first term represents the change of $\zo$ from the variation of
oxygen mass due to star formation and galactic outflows, while the second
term describes the dilution due to gas inflow. We assume the metallicity of
inflowing gas is primordial in this paper.

By further defining the gas consumption timescale
$\tau_*{\equiv}\mgas/\sfr$, we can write out the change rate of oxygen
abundance
\begin{equation}
    \frac{\dd \zo}{\dd t} = \frac{\mocc}{\tau_*}
    - \frac{\zo}{\tau_{\mathrm{dep}}}
    -\left(\frac{\ddot{M}_*}{\sfr} + \frac{\dot{\tau}_*}{\tau_*}\right)\zo,
    \label{eqn:dzodt}
\end{equation}
where
\begin{equation}
    \tau_{\mathrm{dep}}\equiv\tau_*/(1+\eta-r)
\end{equation}
is the gas depletion timescale. For the \powexp{} SFH, we have
\begin{equation}
    \frac{\ddot{M}_*}{\sfr} = \frac{\kappa}{t-t_0} -\frac{1}{\tausfh}.
    \label{eqn:mstardotdot}
\end{equation}
We also assume a scaling relation between $\sfr$ and $\mgas$ similar to
the Schmidt law~\citep{Schmidt1959},
\begin{equation}
\mgas= M_{g,0}\sfr^\epsilon,
    \label{eqn:mgas}
\end{equation}
which provides a good description of the gas reservoir in \eagle{}
galaxies~\citep[as it was put in by hand, see][]{Schaye2008}, so that
\begin{equation}
    \frac{\dot{\tau}_*}{\tau_*} = (\epsilon - 1)\frac{\ddot{M}_*}{\sfr}.
    \label{eqn:dtauovertau}
\end{equation}
Plugging Equation~\ref{eqn:mstardotdot} and \ref{eqn:dtauovertau} into
Equation~\ref{eqn:dzodt}, we arrive at
\begin{equation}
    \frac{\dd \zo}{\dd t} = \left[\frac{\mocc}{\tau_*} -
    \frac{\zo}{\tau_{\mathrm{dep}}}\right] +
    \left[\frac{\epsilon\zo}{\tausfh} -
    \frac{\epsilon\kappa\zo}{t-t_0}\right].  \label{eqn:dzodtpowexp}
\end{equation}
where the term in the first bracket modifies $\mo$ through the combination
of stellar nucleosynthesis, recycling, and outflows, and the second bracket
modifies $\mgas$ through inflows. We refer to the first and second
bracketed terms as ``Outflow+SF'' and ``Inflow'', respectively, in our
later analysis of the conditions for chemical equilibrium.

In addition, Equation~\ref{eqn:dzodtpowexp}, or more generally
Equation~\ref{eqn:dzodtsimple}, serves as the basis for our method of
measuring the strength of outflows from hydrodynamic simulations, as
$\eta(t)$ is the only unknown parameter in the equation that cannot be
measured from simulations in a straightforward manner. We will expand on
the method in detail in~\S\ref{subsubsec:etaceh}.

Setting $\kappa{=}0$ and $\epsilon{=}1$ correspond to the commonly adopted
model of a pure exponential SFH with constant $\tau_*$, reducing
Equation~\ref{eqn:dzodtpowexp} to a simpler form
\begin{equation}
    \frac{\dd \zo}{\dd t} = \frac{\mocc}{\tau_*}
    - \frac{\zo}{\tau_{\mathrm{dep}}}
    + \frac{\zo}{\tausfh}
    = \frac{\mocc}{\tau_*} - \frac{\zo}{\tilde{\tau}},
    \label{eqn:dzodtconst}
\end{equation}
where for the second equality we have adopted the ``harmonic difference
timescale''
\begin{equation}
    \tilde{\tau}\equiv\frac{1}{\tau^{-1}_{\mathrm{dep}}-\tausfh^{-1}},
    \label{eqn:ttilde}
\end{equation}
introduced by \citet{Weinberg2017b}.

\subsubsection{The simple case: constant $\tau_*$ and $\eta$}
\label{subsubsec:simplecem}

Before measuring the time-dependent $\tau_*$ and $\eta$ in the \eagle{}
simulation, we firstly apply our \powexp{} SFH model to the standard CEM
assuming constant values of $\tau_*$ and $\eta$. This simple case serves as
the baseline model against which we compare our comprehensive NE-CEM
in \S\ref{subsec:feature}.

\begin{figure}
    \hspace{-0.5cm}\includegraphics[width=0.48\textwidth]{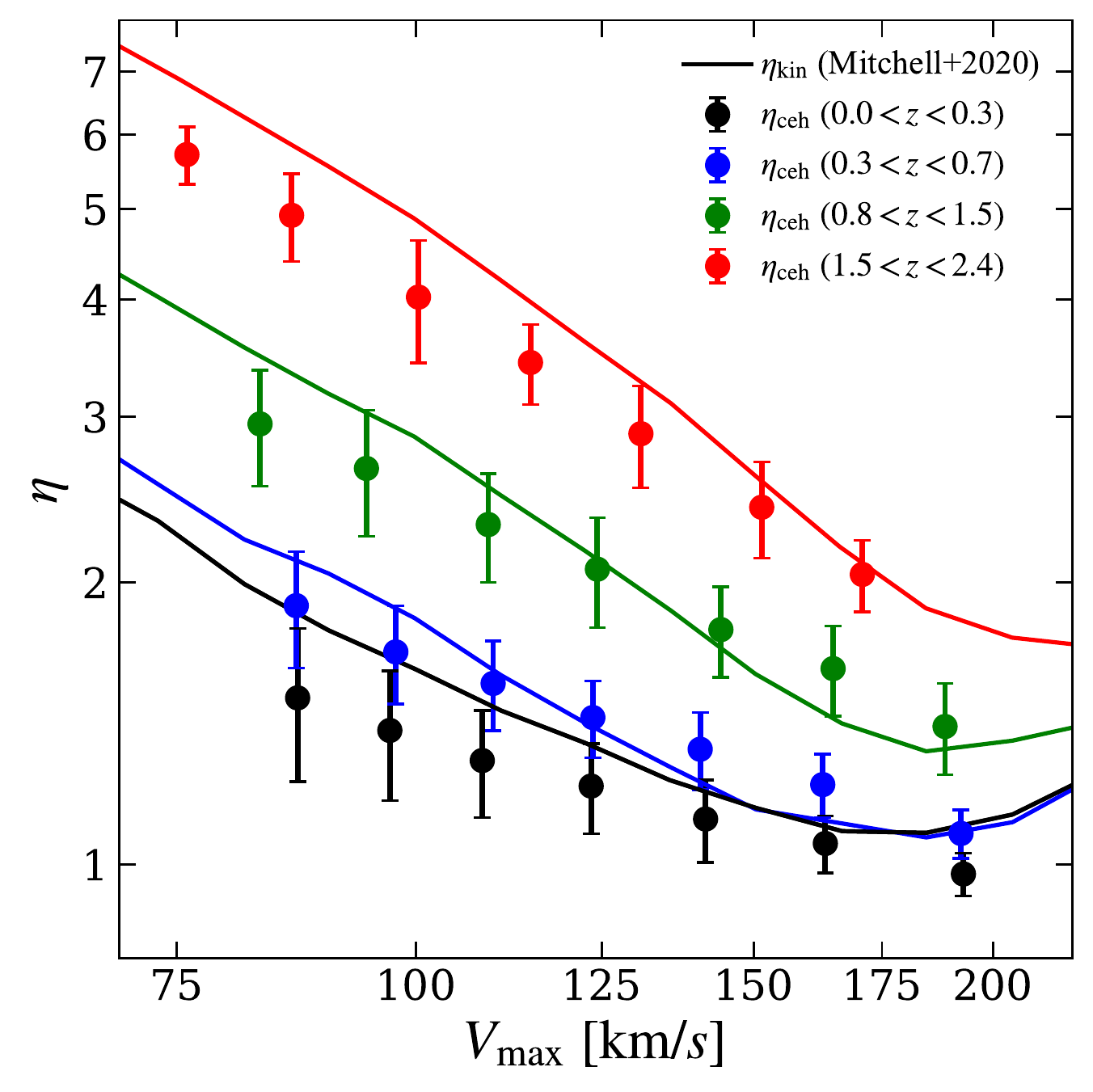}
    \caption{Comparison between the dependences of the mass-loading factor on
	the halo maximum circular velocity measured chemically by our
	method~(circles with errorbars) and kinematically by
	\citet{Mitchell2020}~(solid curves of matching colours) for
	\eagle{} galaxies in four different redshift bins, listed by the
	legend in the top right. The good agreement between the two
	measurements using entirely different methods is quite remarkable.
     \label{fig:eta_comp} }
\end{figure}

Assuming constant $\eta$, we can numerically integrate
Equation~\ref{eqn:dzodt} to obtain the time evolution of $\zo$, and
by further assuming a constant $\tau_*$ by setting $\epsilon{=}1$, the
integration becomes analytic, yielding
\begin{equation}
\zo(t)= \mocc\frac{\tilde{\tau}}{\tau_*}\left(\frac{t}{\tilde{\tau}}\right)^{-\kappa}
    \exp\left({\frac{t}{\tilde{\tau}}}\right)\gamma\left(\kappa+1,\frac{t}{\tilde{\tau}}\right).
\label{eqn:zosimple}
\end{equation}
At $t{\gg}\tilde{\tau}$, this system would approach an equilibrium
metallicity of
\begin{equation}
    \zoeq = \mocc \frac{\tilde{\tau}}{\tau_*},
    \label{eqn:zoeq}
\end{equation}
which is a generic prediction of CEMs with constant $\eta$ and $\tau$ over
an exponential declining tail of star formation~\citep{Weinberg2017}.

Figure~\ref{fig:ceh_simple} shows the average SFHs, CEHs, and the
evolutions of $\dd\zo/\dd t$ in panel (a), (b), and (c), respectively, for
two typical galaxies with the same $\lg\ms^{0.1}{=}9.5$ but different
$\lg\sfr^{0.1}$ of $-0.9$~(red star) and $-0.14$~(blue star), assuming
constant values of $\eta{=}1.5$ and $\tau_*{=}2\,\gyr$.  In panel (a), the
old galaxy~(red curve; $\kappa{=}2.31$, $\tausfh{=}1.96\,\gyr$) has an
early peak in its SFH and is almost quenched by $z{=}0.1$, whereas the
young galaxy~(blue curve; $\kappa{=}{1.2}$, $\tausfh{=}{12.92\,\gyr}$) has
yet to reach peak star formation by $z{=}0.1$. This difference between the
two SFH shapes directly leads to two different trajectories of chemical
evolution in panel (b), where blue and red horizontal dashed lines indicate
the equilibrium metallicities $\lg\zoeq{=}{-1.73}$ and
${-1.97}$~(Equation~\ref{eqn:zoeq}) of the young and old galaxies,
respectively. Meanwhile, blue and red vertical lines indicate the
corresponding equilibrium timescales
 $\tilde{\tau}$ of the young and old galaxies, respectively,
indicating that the young galaxy~($\tilde{\tau}{=}{2.75}\,\gyr$) approaches
chemical equilibrium in a faster pace than its older
counterpart~($\tilde{\tau}{=}{7.92}\,\gyr$). Compared to their lifetimes
until $z{=}0.1$~(12.45 $gyr$), however, the two equilibrium timescales of
both galaxies are rather short.

The details of the CEHs are better illustrated by their differential form
in panel (c), where the solid, dashed, and dotted curves indicate the total
$\dzodt$~(Equation~\ref{eqn:dzodtpowexp}), ``Outflow+SF''~($\mocc/\tau_* -
\zo/\tau_{\mathrm{dep}}$), and ``Inflow''~($\zo/\tausfh -
\kappa\zo/(t-t_0)$) components of $\dzodt$, respectively.  We note that
CEHs in this work refer to the evolution of $\zo$ instead of $\mo$. Both
total change rates of $\zo$ rapidly approach zero, i.e., reaching chemical
equilibrium, after their respective equilibrium timescales $\tilde{\tau}$
indicated by the vertical lines on top.  However, the two galaxies are
enriched in distinct manners. For the young galaxy, the contribution from
the ``Outflow+SF'' component is always positive before
$t_{0.1}{=}12.45\gyr$ and vice versa for the ``Inflow'' component, whereas
for the old galaxy the signs are reversed at large $t$.  The reason is as
follows.  The young galaxy is enriched more slowly~(hence lower $\zo$),
with a longer $\tausfh$, and larger $\kappa$ than the old ones, but the two
galaxies have the same $\tau_*$ and $\tau_{\mathrm{dep}}$, leading to
$\zo^{\mathrm{young}}{<}\mocc\tau_{\mathrm{dep}}/\tau_*{<}
\zo^{\mathrm{old}}$ and
$\tausfh^{\mathrm{young}}\kappa^{\mathrm{young}}>t>\tausfh^{\mathrm{old}}\kappa^{\mathrm{old}}$
simultaneously when $t_1 > t \gg t_0$. In essence, for the young galaxy,
metal production overwhelms metal ejection due to the overall low level of
enrichment in the ISM, while metal dilution is effective due to the ample
inflow of pristine gas. The situation for the old galaxy is reversed ---
metal production due to waning star formation cannot keep up with the metal
loss due to outflows loaded with highly-enriched ISM, while metal dilution
turns into metal concentration as the meager inflow cannot catch up with
the rate at which the gas is consumed.  At $t{\gg}t_1$, however, the
``Outflow+SF'' term of young galaxy will cross zero and becomes negative as
$\zo$ keeps increasing, while the ``Inflow'' term instead becomes positive,
leading to similar trajectories as the old galaxy.  Eventually, both
components conspire to reach equilibrium metallicity $\zoeq$, following the
exponentially declining $\sfr$ regardless of their early SFHs.

Finally, panel (d) of Figure~\ref{fig:ceh_simple} demonstrates that the
value of $\eta$ can be solved from the combination of SFH~(top left) and
CEH~(top right) using Equation~\ref{eqn:dzodtpowexp}~(bottom left). Red and
blue circles mark two random epochs of $2.5 \,\gyr$ and $8.5 \,\gyr$ for
the old and young galaxies, respectively~(also shown across the other three
panels of Figure~\ref{fig:ceh_simple}).  Solid red~(blue) line shows the
variation of $\dzodt$ as a function of $\eta$ at $2.5\,\gyr$~($8.5\,\gyr$)
for the old~(young) galaxy, given by Equation~\ref{eqn:dzodtpowexp}.
Horizontal dashed lines indicate the true values of $\dzodt$ measured from
the CEHs~(horizontally aligned with the circles in panel (c)). Therefore,
the intersections between the solid and dashed lines of the same colour
yield the correct values of $\eta$~(i.e., the x-axis value of the circles,
$\eta{=}1.5$) that are required by the consistency between the SFH and CEH
of the same galaxy at that particular epoch. Next we will apply this method
to the \eagle{} simulation to reconstruct the MLHs of outflows in \eagle{}.

\subsubsection{Modelling mass-loading histories $\eta(t)$ in \eagle{}}
\label{subsubsec:etaceh}

To explore robust modelling of $\eta$ that can describe the galactic winds
in the \eagle{} simulation and in the observations, we first need to
reconstruct the average MLHs of \eagle{} galaxies from their CEHs. For this
purpose, we develop a novel method to accurately measure $\eta$ chemically
and describe the method in detail below. Although developed with the
\eagle{} simulation in mind, the method can be easily applied to any
hydrodynamical simulation that explicitly tracks oxygen in the ISM.

\begin{figure*}
\centering \includegraphics[width=0.9\textwidth]{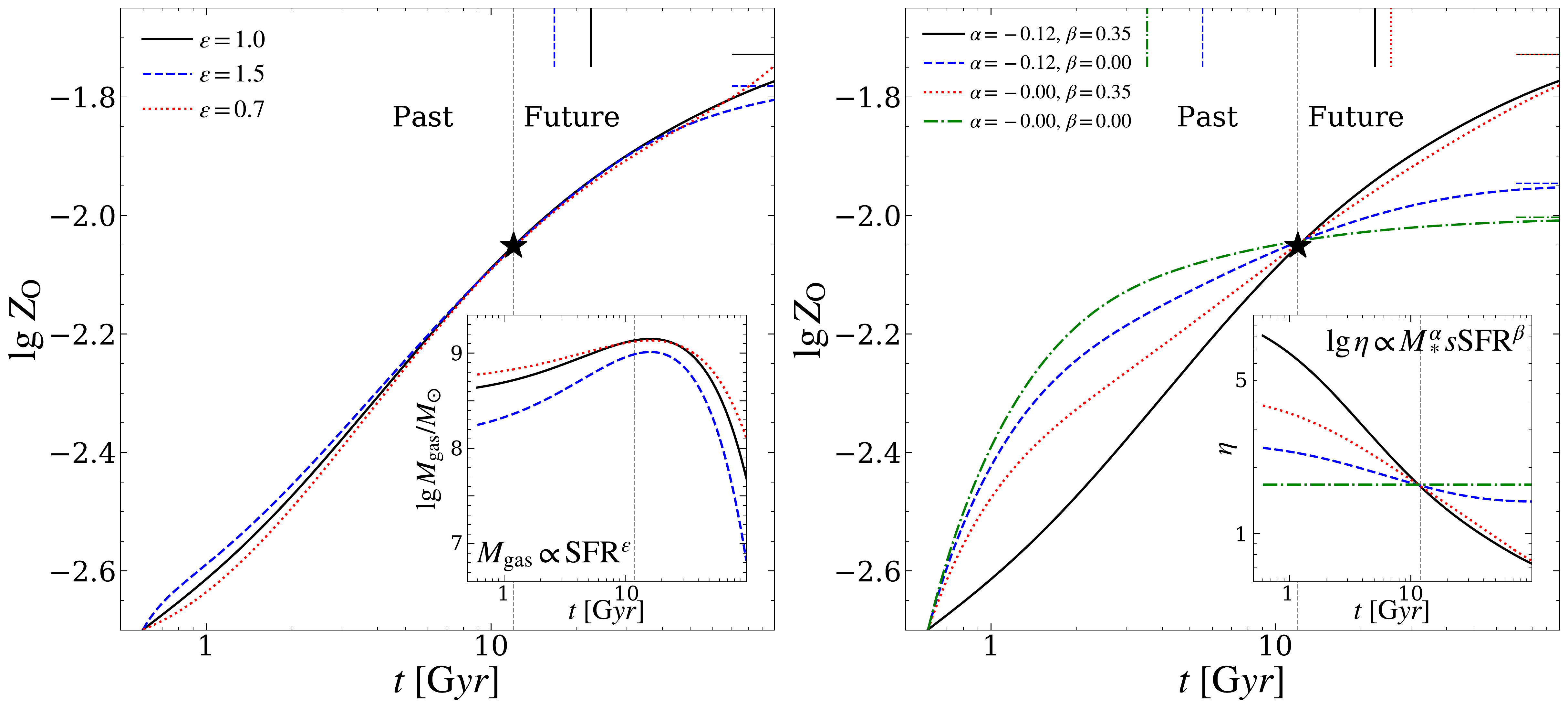}
    \caption{Impact of $\epsilon$~(left panel) and the combination of
    $\alpha$ and $\beta$~(right panel) on the chemical evolution of a
    typical galaxy from birth to $t{=}100\,\gyr$.  The SFH of the galaxy is
    the same as that of the ``young'' galaxy in the panel (a) of
    Figure~\ref{fig:ceh_simple}.  In each panel, we adjust the
    normalizations of $\mgas$~(left) or $\eta$~(right) so that the
    trajectories of chemical enrichment all pass through the same value of
    $\zo$ today at $t{=}12.45\,\gyr$. In the left panel, black solid, blue
    dashed, and red dotted curves indicate the different chemical
    enrichment trajectories assuming $\epsilon{=}1$~(i.e., constant
    $\tau_*$), $1.5$, and $0.7$, respectively.  The inset panel on the left
    shows the corresponding evolutions of $\mgas$. In the right panel,
    black solid, blue dashed, red dotted, and green dot-dashed curves
    indicate the predicted chemical enrichment from $(\alpha,
    \beta){=}(-0.12, 0.35)$, $(-0.12, 0)$,  $(0, 0.35)$, and $(0,
    0)$~(i.e., constant $\eta$), respectively.  Horizontal and vertical
    short lines in each panel indicate the corresponding equilibrium
    metallicities and timescales of the CEHs~(except for $\epsilon{=}0.7$
    in which case the galaxy could never reach chemical equilibrium).
    \label{fig:model_comp} }
\end{figure*}

For each of the 28 bins of fixed $\ms^{0.1}$ and $\sfr^{0.1}$ in
Figure~\ref{fig:sfheagle}, we firstly measure the sum of the star-forming
gas mass $\mgas(t)$~(\texttt{MassType\_Gas}) and the sum of oxygen mass in
the star-forming gas
$\mo(t)$~(\texttt{SF\_Mass}$\times$\texttt{SF\_Oxygen}) from all the
progenitors at each of the 14 outputs between $z{=}2.24$ and $z{=}0.1$,
using the merger trees built in~\S\ref{subsubsec:mtree}.  Secondly, from the
measured $\mgas(t)$ and $\mo(t)$ we can compute $\zo(t)$ as well as its
derivative $\dzodt(t)$ from the smooth CEH derived from quadratically
interpolating $\zo$ over the 14 snapshots.  Lastly, we solve for $\eta(t)$
by applying the measured quantities to Equations~\ref{eqn:modot} and
\ref{eqn:dzodtsimple}, following the methodology illustrated in the panel
(d) of Figure~\ref{fig:ceh_simple}.  Similar to the practice in the SFH
measurements, we employ the ``collapsed'' tree to compute an effective
mass-loading factor for the synthetic progenitor in each epoch.  To avoid
confusion when comparing with the $\eta$ measured by other techniques
in~\S\ref{subsubsec:mitchell}, we refer to our mass-loading factors derived
from the CEHs as $\etaceh$. In total, we have obtained $\etaceh$
measurements for $14\times28{=}392$ average galaxies, i.e., 14 epochs for
each of the 28 bins of galaxies at fixed $\ms^{0.1}$ and $\sfr^{0.1}$.

As discussed in the Introduction, the mass-loading factor of a galaxy
should depend on the galaxy internal properties, and its apparent variation
with redshift is due to the redshift evolution of those internal
properties. The most important among them are the stellar mass $\ms$, which
determines the strength of the gravitational potential, and the specific
star formation rate $\ssfr$, which is associated with the specific
injection rate of energy and/or momentum that drive the galactic winds.
Therefore, we build a simple empirical model of $\etaceh(t)$ by
parametrising $\lg\etaceh(t)$ as the product of two power-laws of $\ms(t)$
and $\ssfr(t)$,
\begin{equation}
\lg \left(\frac{\eta}{\eta_0}\right) = f \left(\frac{\ms}{M_{*,0}}\right)^\alpha
\left(\frac{\ssfr}{\ssfr_0}\right)^\beta,
    \label{eqn:etacehmodel}
\end{equation}
where $f$ is the overall normalization parameter, and we fix the three
pivot values to $\lg\eta_0{=}{-}0.2$, $\lg M_{*,0}{=}9.5$, and
$\lg\ssfr{=}{-}9.5$ throughout this paper.  Note that the parameterisation
in Equation~\ref{eqn:etacehmodel} implies that $\eta$ is always greater
than $\eta_0{=}0.631$, which should be comfortably lower than the
mass-loading factors in star-forming galaxies
below a few ${\times}10^{10}\msol$, and even more so considering the
outflows are likely more enriched than the ISM~(i.e., $\zeta$ is always
great than $\eta$).

The two slopes of the power-laws in Equation~\ref{eqn:etacehmodel} are the
key parameters that we aim to constrain from observations in this paper.
In particular, $\alpha$ is the slope of (log) mass-loading dependence on
stellar mass, where simple theoretical arguments expect
$\eta{\propto}\ms^{{-}1/3}$ for momentum-conserving
winds~\citep{Murray2005} and $\eta{\propto}\ms^{{-}2/3}$ for
energy-conserving winds~\citep{CC85, Heckman1990}; $\beta$ is the slope of
the $\ssfr$-dependence. For stochastic explosion of isolated massive stars
in the low-$\ssfr$ regime, the outflows are highly inefficient with low
$\eta$ because most supernova remnants radiate significant energy away
before breaking out of the galactic discs~\citep{Koo1992, Murray2011}.
However, in starburst systems the stellar explosions are temporally
correlated and spatially clustered~\citep{Gentry2017, Yadav2017,
Gentry2019}, thereby driving superbubbles that rapidly break out of the
discs and power strong galactic outflows with high $\eta$~(\citep{Kim2017,
Yadav2017, Vasiliev2017, Fielding2018}. Roughly speaking, the measurement
of $\alpha$ provides useful constraints on the wind driving mechanism,
whereas the $\beta$ measurement could probe the impact of clustered
supernovae on stellar feedback.

\begin{figure*}
\centering \includegraphics[width=0.8\textwidth]{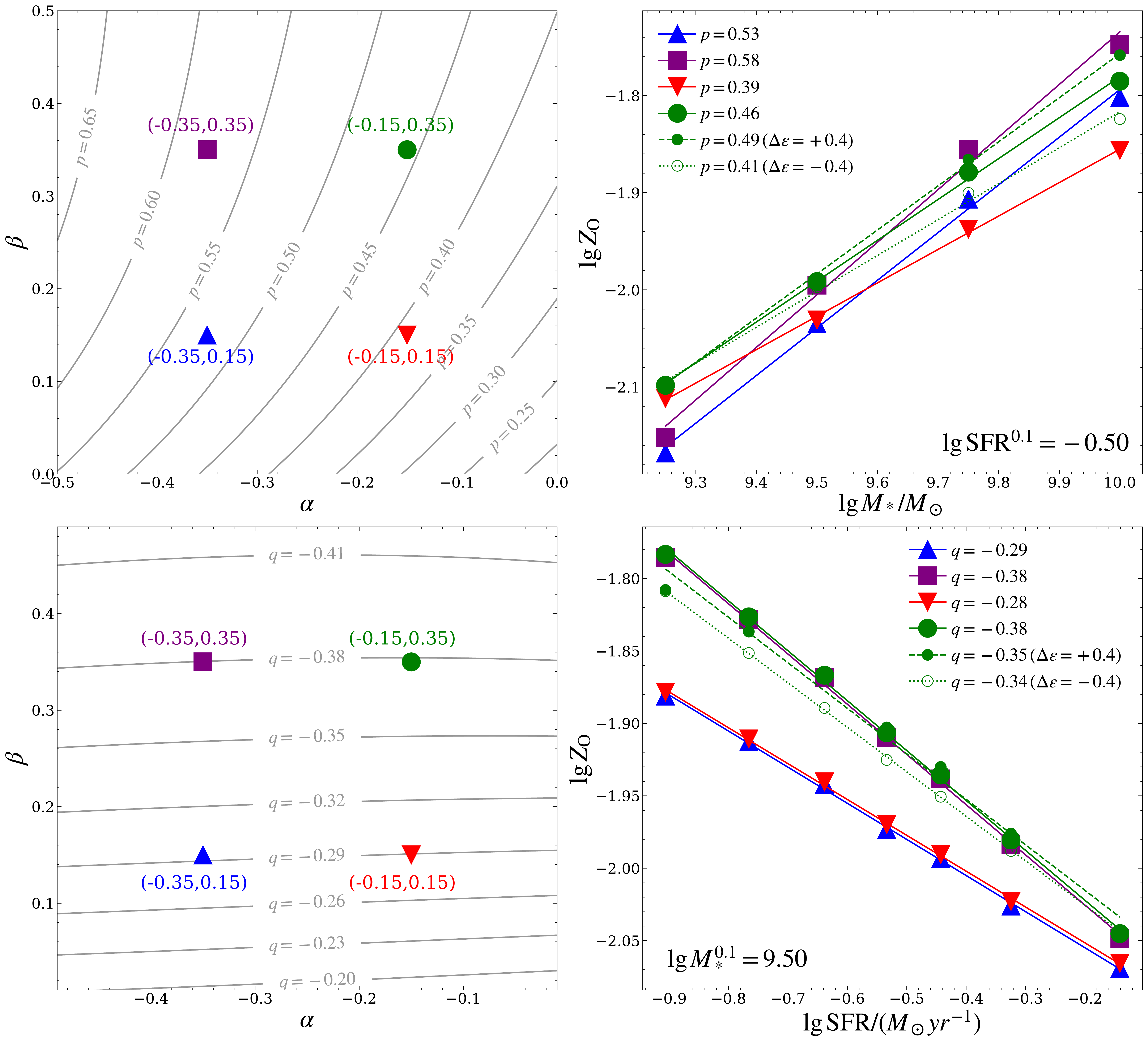}
    \caption{Connection between the slope parameters of $\eta$~($\alpha$ and
    $\beta$) and the slope of the $\zo$-$\ms$ relation at fixed SFR~($p$;
    top row) or the slope of the $\zo$-$\mathrm{SFR}$ relation at fixed
    $\ms$~($q$; bottom row), predicted by the fiducial NE-CEM for \eagle{}
    galaxies at $z{=}0.1$.  In each row, the left panel shows the iso-$p$
    contours predicted by the fiducial NE-CEM on the 2D plane of $\alpha$
    vs. $\beta$ with all other parameters fixed, while the right panel
    shows the predicted $\zo$-$\ms$ relations at
    $\lg\mathrm{SFR}{=}{-}0.5$~(top) or the predicted $\zo$-$\mathrm{SFR}$
    relations at $\lg\ms{=}9.5$~(bottom) for four different combinations of
    $\alpha$ and $\beta$~(solid lines through large symbols) indicated by
    the four coloured symbols in the left panel~(with the parameter values
    displayed inside the parentheses). The measured values of the
    slopes are indicated by the legends on the top left. In addition,
    dashed and dotted lines through the small green filled and open circles
    in the right panel indicate the predicted relations after perturbing
    the fiducial value of $\epsilon{=}0.91$ by ${+}0.4$ and ${-}0.4$,
    respectively.  \label{fig:slope}}
\end{figure*}

We perform a multiple linear regression analysis using
Equation~\ref{eqn:etacehmodel} over the 392 sets of $\etaceh$, $\ms$, and
$\ssfr$ measurements~(assuming equal weights), yielding the best-fitting
values of $f{=}0.45{\pm}{0.01}$, $\alpha{=}{-}0.15{\pm}{0.02}$, and
$\beta{=}0.34{\pm}{0.02}$.  Figure~\ref{fig:etafit} compares the
predictions~(x-axis) from the best-fitting model of
Equation~\ref{eqn:etacehmodel}~(indicated in the top left corner) to the
$\etaceh$ directly inferred from the CEHs~(y-axis) for the 392 systems in
\eagle{}~(circles). The colour and size of each circle represent the
$\ssfr$ and $\ms$ of the galaxy indicated by the colourbar and legend in
the bottom right corner, respectively.  The circles are mostly aligned with
the black diagonal line~(i.e., the one-to-one relation), showing that
Equation~\ref{eqn:etacehmodel} is a good model for describing the
mass-loading factors in the \eagle{} simulation at all redshifts.

Figure~\ref{fig:ceheagleall} demonstrates the efficacy of our best-fitting
model of $\etaceh$ when combined with the \powexp{} SFH in predicting the
CEHs of \eagle{} galaxies.  The four panels compare the CEHs directly
measured from $\eagle${}~(circles) with those predicted by the combined
model of the best-fitting $\etaceh$ from Equation~\ref{eqn:etacehmodel} and
the \powexp{} SFHs, for the same sets of galaxy samples defined in
Figure~\ref{fig:sfheagleall}.  In each inset panel, we show the oxygen
abundance at $z{=}0.1$ as a function of $\sfr^{0.1}$ directly measured from
the simulation~(stars), which is successfully predicted by the best-fitting
combined model~(black solid curve). Overall, the predicted CEHs are in good
agreements with the direct measurements, especially at $z{=}0.1$ where the
$\zo(\ms, \sfr)$ relation is an observable in the mock test of
\S\ref{subsec:mock}.

\subsubsection{Comparison with \citet{Mitchell2020}$: \etaceh$ vs. $\etakin$}
\label{subsubsec:mitchell}

To further validate our reconstruction of the MLHs of outflows, we can
compare our chemically-inferred $\etaceh$ with the mass-loading factors
directly measured from counting the wind particles in the \eagle{}
simulation. However, direct measurement of $\eta$ in hydrodynamic
simulations is never straightforward, primarily because $\eta$ is often not
a direct parameter put in by hand~\citep[though could be set at injection;
see e.g., ][]{Pillepich2017}, but requires a careful but somewhat arbitrary
identification and tracking of wind particles. In particular, to measure
the mass loading factors in the same \eagle{} simulation employed by this
work, \citet{Mitchell2020} firstly identified the ISM particles by
including the star-forming gas and some dense portion of the
non-star-forming gas, and then from those ISM particles selected galactic
wind particles as those with time-averaged radial velocities higher than
one quarter of the maximum circular velocity of the halo $\vmax$ and
instantaneous radial velocities above $\vmax/8$. We hereafter refer to the
mass-loading factors measured by \citet{Mitchell2020} using this kinematic
criteria as $\etakin$, as opposed to $\etaceh$ we measured chemically in
\S\ref{subsubsec:etaceh}.

\begin{figure*}
\centering\includegraphics[width=0.8\textwidth]{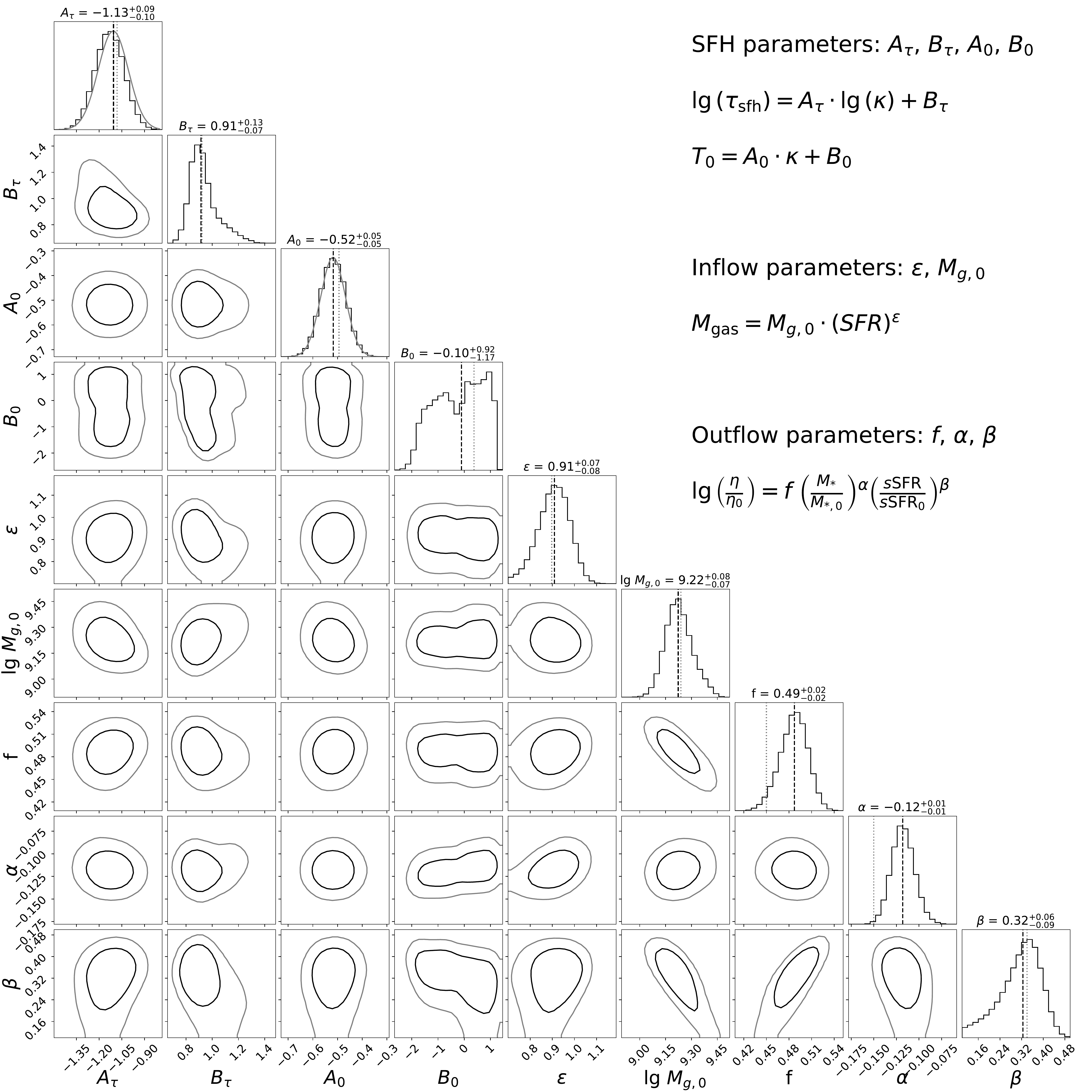}
    \caption{Constraints on the NE-CEM parameters from the mock test of using
        the $\zgas(\ms, \sfr)$ relation of \eagle{} galaxies at $z{=}0.1$
        as input data in our Bayesian inference.  Contours in the
        off-diagonal panels indicate the 68\%~(black) and 95\%~(gray)
	confidence regions on the 2D plane that comprised of all the pair
	sets of nine model parameters~(parameter definitions shown in the
	top right formulae). Histograms in the diagonal panels show the 1D
	posterior distributions of individual parameters, with gray
	distributions in the $A_{\tau}$ and $A_0$ panels representing the
	Gaussian priors.  Dark dashed and
	light dotted vertical lines in each diagonal panel indicate the
	parameter value estimated from the 1D posterior mean and that
	measured directly from the simulation, respectively. The $1\sigma$
	posterior constraint of each parameter is listed on top of the
	respective diagonal panel.  \label{fig:mcmc_eagle} }
\end{figure*}

Figure~\ref{fig:eta_comp} compares our chemically-measured
$\etaceh$~(circles with errorbars) with the kinematically-measured
$\etakin$ by \cite{Mitchell2020}~(solid curves) in four different redshift
bins between $z{=}0$ and $z{=}2.4$. The redshift binning exactly follows
that in the Fig. 3 of \cite{Mitchell2020}. To facilitate comparison, we
also bin our galaxies by the $\vmax$ of their host haloes, and predict
$\etaceh$ from their $\ms$ and $\ssfr$ using the best-fitting model shown
in Figure~\ref{fig:etafit}. The errorbars are the $1{-}\sigma$ scatters
about the mean. The $\etakin$ curves all have inflections at
$\vmax{>}200\kms$, signalling the takeover by AGN feedbacks in more massive
galaxies.  Below $\vmax{=}200\kms$ where stellar feedback dominates, the
two types of measurements are generally consistent with each other within
the errorbars across all four redshift bins, except for some of the
low-mass systems in the highest redshift bin~($\vmax{<}100$). Although the
agreement is not perfect, the fact that two entirely different methods, one
chemical and the other kinematical, yield mass loading factors that are
quantitatively similar, is quite remarkable.

The overall agreement shown in Figure~\ref{fig:eta_comp} is also very
encouraging --- our method of parametrising $\etaceh$ as a function of
$\ms$ and $\ssfr$ in Equation~\ref{eqn:etacehmodel} not only appears
reasonable from a theoretical perspective, but also predicts the correct
scaling of the mass-loading factor with $\vmax$ and $z$. In particular,
reproducing the correct redshift evolution of $\eta$ is highly nontrivial.
As we emphasized in~\S\ref{subsubsec:etaceh}, $\eta$ should in principle be
determined by the internal state of a galaxy, whose redshift evolution then
drives the apparent dependence of $\eta$ on $z$. Therefore, this agreement
seen in Figure~\ref{fig:eta_comp} further strengthens our belief in the
physical robustness behind our parametrisation of $\eta$, especially when
applied to the real observations.

\section{A comprehensive non-equilibrium chemical evolution model}
\label{sec:necem}

\subsection{Assembling the NE-CEM}
\label{subsec:assembly}

After developing analytic models for the histories of star formation and
mass-loading in the \eagle{} simulation, we now assemble a comprehensive
NE-CEM for the star-formation main sequence observed at $\zobs$ as follows.
For an observed $\zo(\msobs, \sfrobs)$ relation consisting of $N$ galaxy samples
within different 2D bins of ($\msobs$, $\sfrobs$),
\begin{itemize}
    \item We describe the average SFH of each sample using
    Equation~\ref{eqn:sfhalt}.  For each sample, we solve the value of
    $\kappa$  from $\msobs$ and $\sfrobs$
	via the inverse function of $\mathcal{F}(\kappa)$ in
	Equation~\ref{eqn:kappamap}, and then compute
    $\tausfh$ and $t_0$ from $\kappa$ using Equations~\ref{eqn:taukappa}
    and \ref{eqn:t0kappa}, respectively. Therefore, the $N$ average SFHs
    can be predicted with just four parameters $\{A_{\tau}, B_{\tau}, A_0,
    B_0\}$.

    \item The gas mass $\mgas(t)$ is predicted from SFHs using
    Equation~\ref{eqn:mgas}, which requires two parameters
	$\{M_{g,0}\,\epsilon\}$.  Note the gas mass $\mgas$ in our NE-CEM
	model is a measure of the effective amount of ISM that has
	participated in the recycling, mixing, and loading of metal under
	our one-zone open-box assumption. Therefore, although
	Equation~\ref{eqn:mgas} resembles the observed volumetric
	star-formation law, we do not expect the slope $\epsilon$ to be
	directly linked to that of the observed star-formation law.

    \item The mass-loading factors $\eta(t)$ are predicted from $\ms(t)$
    and $\ssfr(t)$ from Equation~\ref{eqn:etacehmodel} with three
    parameters $\{f, \alpha, \beta\}$. Finally, the evolution of $\zo$ can
	be predicted by numerically integrating
	Equation~\ref{eqn:dzodtpowexp} for each of the $N$ galaxy samples
	from $t_0$ to $\tobs$.
\end{itemize}
Despite the comprehensiveness of the analytic framework, our full NE-CEM
has only nine parameters in total:
$\{A_\tau,B_\tau,A_0,B_0,\epsilon,M_{g,0},f,\alpha,\beta\}$.  Among the
three steps, the first step can be significantly improved in the future by
applying SED fits to the average spectra assuming \powexp{} SFH~(Chen et
al. {\it in prep}).

\subsection{NE-CEM: features of time-varying $\tau_*$ and $\eta$}
\label{subsec:feature}

Figure~\ref{fig:model_comp} illustrates the impact of different
$\epsilon$~(left panel) and combinations of $(\alpha, \beta)$~(right panel)
on the trajectory of chemical enrichment for a typical \powexp{} SFH~(same
as that of the ``young'' galaxy shown in Figure~\ref{fig:ceh_simple}) under
our NE-CEM framework.  Unsurprisingly, compared to the simple case with
constant $\tau_*$ and $\eta$ shown in Figure~\ref{fig:ceh_simple}, the full
NE-CEM with time-varying $\tau_*$ and $\eta$ exhibits significantly more
complex behaviors. In the left panel, we vary the value of $\epsilon$ to be
$1$~(black solid), $1.5$~(blue dashed), and $0.7$~(red dotted) while
keeping other parameters fixed except for $f$, which we adjust to make the
three trajectories arrive at the same $\zo$ at $z{=}0$~(star symbol).  The
equilibrium metallicities and timescales are indicated by the
corresponding short horizontal and vertical lines, respectively. The
equilibrium timescale is defined as the epoch at which $\zo$ reaches
$(1-1/e)=63.2\%$ that of the equilibrium value. We note that the
equilibrium timescales in the NE-CEM differ from $\tilde{\tau}$~(previously
defined in Equation~\ref{eqn:ttilde}), which is only valid when both
$\tau_*$ and $\eta$ stay constant. For NE-CEM, the equilibrium timescale is
determined by the time evolution of $\eta$~(and also $\tau_*$, but to a
lesser degree).  The system with $\epsilon{=}0.7$ does not have an
equilibrium metallicity because
$\dzodt{\rightarrow}\epsilon\zo/\tausfh{>}0$~(Equation~\ref{eqn:dzodtpowexp})
at large $t$.  The inset panel shows the respective evolutions of
$\mgas(t)$, with the $\epsilon{=}1$ curve having exactly the same shape as
the SFH because the $\mgas$-$\sfr$ scaling is linear. Note that the gas
reservoir $\mgas$ is not ``steady'' and varies more than an order of
magnitude across the lifetime of the galaxy in all three cases.  Similarly,
the right panel shows the impact on the CEH caused by the different MLHs
due to the change in $\alpha$ and/or $\beta$~(indicated by the legend in
the top left corner), with the inset panel illustrating the variations of
$\eta$ as functions of time.

\begin{figure}
    \includegraphics[width=0.48\textwidth]{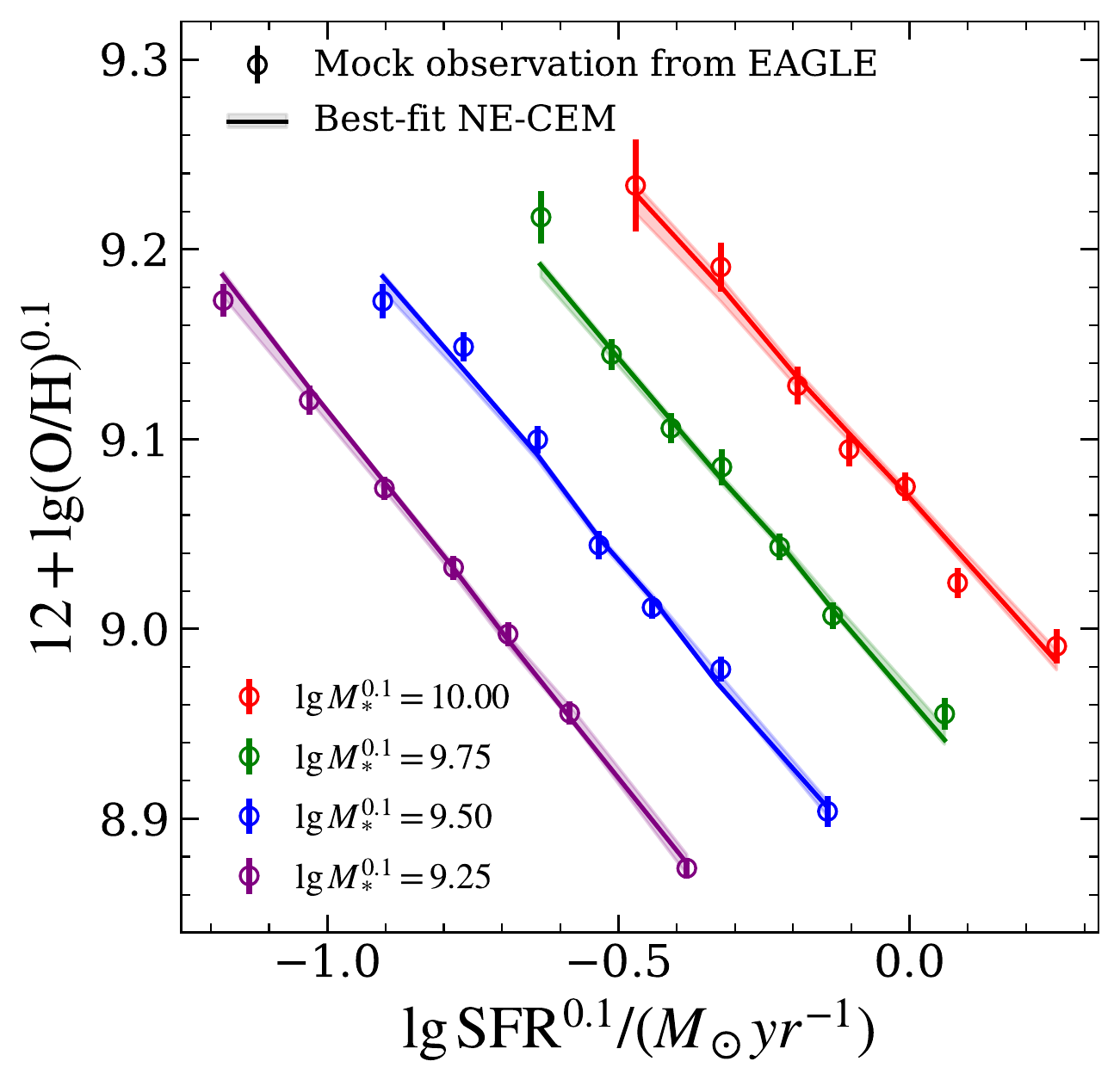}
    \caption{Comparison between the $\zgas(\ms, \sfr)$ relation at
    $z{=}0.1$ measured directly from \eagle{}~(open circles with errorbars)
    and predicted by the posterior mean NE-CEM model~(curves with $1\sigma$
    uncertainty bands). The errorbars on the data points are the standard
    errors of the mean, while the four colours indicate the four stellar
    mass bins listed in the bottom left corner.  \label{fig:obs_eagle} }
\end{figure}

Comparing the two panels of Figure~\ref{fig:model_comp}, we find that the
impact of $\epsilon$ on the CEH is relatively small compared to the that
caused by $\alpha$ and $\beta$. In particular, the fiducial
$\eta(t)$~($\alpha{=}-0.12$, $\beta{=}{0.35}$; black solid curve on the
right panel) measured from the \eagle{} simulation results in a large
equilibrium timescale of $22.4\,\gyr$, with the metallicity steeply
increasing at the current epoch; The CEH predicted using $\alpha{=}0$ and
$\beta{=}{0.35}$~(red dotted) experiences a steeper initial rise but a
shallower late surge than the fiducial curve, yielding a slightly larger
equilibrium timescale of $25.5\,\gyr$.  Thus, both systems with positive
$\beta$~(hence significant time-variation of $\eta$; inset panel) would
spend their past lifetime in an non-equilibrium state, which can only be
accurately described by an NE-CEM.  In contrast, the other two cases with
$\beta{=}0$~(blue dashed) and constant $\eta$~(green dot-dashed) experience
rapid enrichment at early times and approach chemical equilibrium at
$5.6\,\gyr$ and $3.5\,\gyr$, respectively, well before the
current epoch, and therefore can be reasonably described by the standard
equilibrium-type CEMs. Comparing the trajectories of the CEH with that of
the MLH~(inset panel), we can see that the shapes of the histories are
strongly correlated, suggesting a potentially tight connection between the
slopes of the observed $\zgas(\ms, \sfr)$ relation and the slopes of
mass-loading --- $\alpha$ and $\beta$.

Note the asymptotic behaviors at $t{\rightarrow}{+}\infty$ shown in
Figure~\ref{fig:model_comp} depend sensitively on the tail of the SFH. In
the real Universe the galaxies would likely be quenched~\citep[e.g., also
by outflows;][]{Garling2022} at some large but finite $t$, i.e., having
$\sfr{=}0$ and $\mgas{=}0$, instead of continuing forming stars with an
infinitesimal rate. For describing these ``red and dead'' galaxies, in the
future we plan to add a transition from the \powexp{} model to a linear
ramp after some quenching timescale, as proposed by \citet{Simha2014}.

\subsection{Connecting $\zo(\ms, \sfr)$ to $\etaceh(\ms, \ssfr)$}
\label{subsec:slopes}

\begin{figure*}
\centering\includegraphics[width=0.8\textwidth]{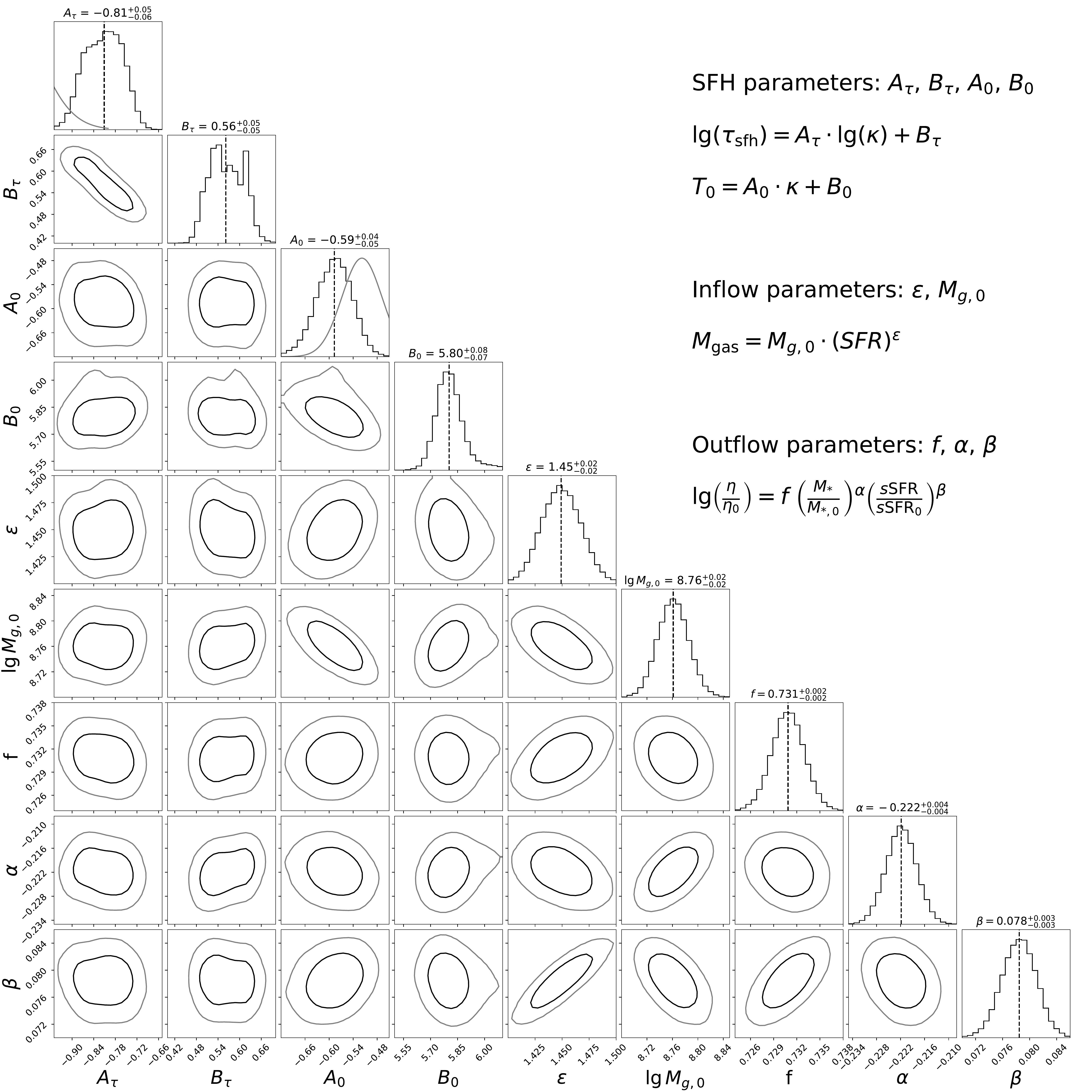}
    \caption{Similar to Fig~\ref{fig:mcmc_eagle}, but for the NE-CEM
    constraints using the $\zgas(\msobs, \sfrobs)$ relation measured from
    SDSS by \citet{Mannucci2010}.  \label{fig:mcmc_sdss} }
\end{figure*}

Following the discussion in \S\ref{subsec:feature}, we discover that the
average CEH of a galaxy sample depends sensitively on the shape of the
mass-loading history, which in turn is determined by the combination of
$\alpha$ and $\beta$ for any given SFH. Therefore, it is of vital importance to
ask the following question: how are the slopes of the observed $\zo(\ms,
\sfr)$ relation, i.e., slope $p$ of the $\zo{-}\ms$ relation at fixed
$\sfr$ and slope $q$ of the $\zo{-}\sfr$ relation at fixed $\ms$, connected
to $\alpha$ and $\beta$ in Equation~\ref{eqn:etacehmodel}?  The key to
unlocking the underlying physics of galactic outflows from FMR observations
lies in the answer to this question. In the simplest~(but unphysical)
scenario, if a galaxy is always in chemical equilibrium, by reaching
$\zoeq$ instantaneously~(i.e., $\tilde{\tau}{\ll}\tau_*$) at every epoch
with changing $\ms$ and $\ssfr$~(hence changing $\eta$), its metallicity at
that epoch should be connected to its instantaneous mass-loading factor
$\eta(\ms, \ssfr)$ via
\begin{equation}
    \zo(\ms, \sfr) \simeq \frac{\mocc}{1+\eta(\ms,\ssfr)-r} \sim \frac{\mocc}{\eta(\ms,\ssfr)},
    \label{eqn:pqnaive}
\end{equation}
from which we expect $p$ and $q$ to depend solely on
$\alpha{-}\beta$ and $\beta$, respectively.

To investigate the connection between $(p, q)$ and $(\alpha, \beta)$ in
NE-CEM, we set up an experiment by computing the values of $p$ and $q$ at
$\lg\sfr^{0.1}{=}-0.5$ and $\lg\ms^{0.1}{=}9.5$, respectively, on a grid of
$(\alpha, \beta)$ using the fiducial model calibrated against the \eagle{}
simulation while keeping the seven other parameters fixed.  The result of
this experiment is displayed in Figure~\ref{fig:slope}, where we show the
contours of $p$ and $q$~(contour levels shown in-line) on the $\alpha$ vs.
$\beta$ plane in the top left and bottom left panels, respectively.  The
four different coloured symbols in each contour panel indicate the loci of
the four sets of $(\alpha,\beta)$~(values in the parentheses by each
symbol) that we select to show the predictions of the scaling relations~(in
matching colours and symbols) on the corresponding right panel.

In the top left panel of Figure~\ref{fig:slope}, the contour lines of
constant $p$ are largely aligned in the diagonal direction, i.e., lines of
constant $\alpha{-}\beta$, when $|\alpha|$ and $\beta$ are both small.
This is consistent with the naive expectation from
Equation~\ref{eqn:pqnaive}, suggesting that outflows are the primary driver
of $\zo$ whether it be in or out of equilibrium. In the top right panel,
however, although the predicted relations~(best-fitting solid lines through
large symbols) have slopes increasing from $p{=}0.39$~(red inverted
triangle) to $0.58$~(purple square) as $\alpha{-}\beta$ decreases from
${-}0.3$ to ${-}0.7$, they differ substantially between $p{=}0.46$~(green
circles) and $p{=}0.53$~(blue triangle) despite having the same
$\alpha{-}\beta{=}{-}0.5$.  In addition, the value of $p$ can be modified
by changing $\epsilon$. The two small green circles show the predicted
scaling after we increase(dashed line through filled circles) or
decrease~(dotted line through open circles) the value of $\epsilon$ by
$0.4$ while keeping all other parameters the same as those for the large
green circle. Consequently, the value of $p$ increases by $0.03$ and
decreases by $0.05$, respectively. Therefore, the slope in the
mass-metallicity relation is largely determined by the stellar
mass-dependence of $\eta$ in the NE-CEM, but is also strongly affected by
the $\ssfr$-dependence of $\eta$ as well as $\epsilon$, i.e., the
$\sfr$-dependence of $\tau_*$.

\begin{figure*}
\centering\includegraphics[width=0.96\textwidth]{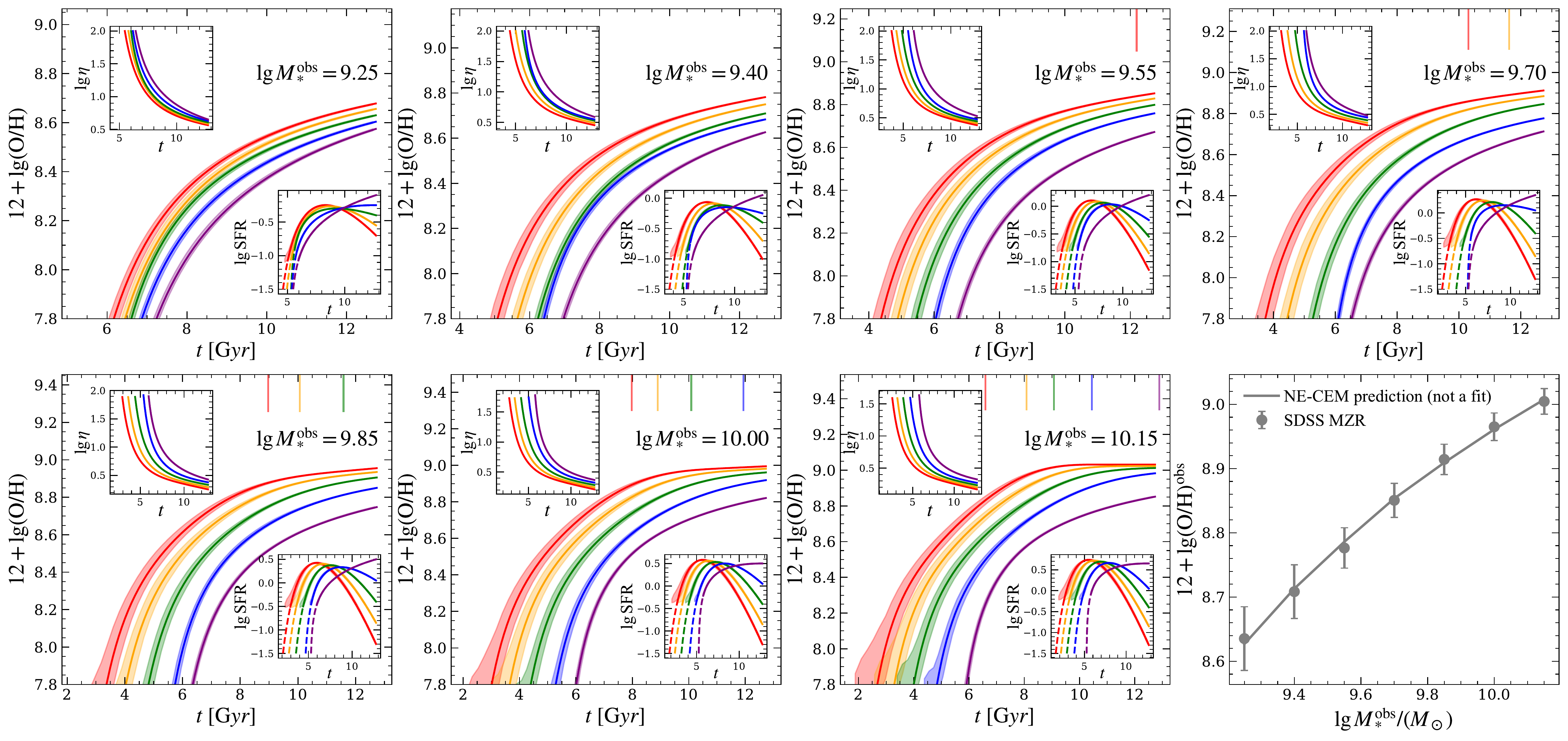}
    \caption{Histories of chemical enrichment~(main panels), mass
    loading~(inset panels in the top left corners), and star
    formation~(inset panels in the bottom right corners) predicted by the
    posterior mean NE-CEM for SDSS galaxies in seven stellar mass
    bins~(with $\msobs$ increasing from left to right, top to bottom). The
    bottom rightmost panel shows the average metallicities of the seven
    $\msobs$ bins measured from SDSS~(gray circles withe errorbars) and
    predicted by the posterior mean NE-CEM~(gray curve; not a fit). In each
    panel of the seven stellar mass bins, we show the predicted CEHs, MLHs,
    and SFHs along with their respective $1\sigma$ uncertainty bands for
    galaxies in five $\sfrobs$ bins. To avoid clutter, we do not show all
    the $\sfrobs$ bins for the high-$\msobs$ samples. The equilibrium
    timescale of each CEH is indicated by the short vertical line on the
    top with matching colour, should it be shorter than the age of the
    Universe today~(increasingly more common with increasing $\msobs$).
    The dashed portions of the SFHs in the inset panels represent the
    extrapolation before the starting
    point of each SFH~(i.e., epoch at which galaxies accumulated 1\% of the
    final observed mass).  \label{fig:eh_sdss} }
\end{figure*}

By the same token, the contour lines of constant-$q$ in the bottom left
panel of Figure~\ref{fig:slope} are mostly horizontal, suggesting the slope
of the $\zo$-$\sfr$ relation is primarily set by $\beta$, modulo some
residual dependence on $\alpha$. Similar to the top right panel, the
predicted slopes in the bottom right panel confirm the suggestion, with
little discrepancy between blue triangles~($-0.29$) and red inverted
triangles~(-0.28), nor between green circles~(-0.38) and purple
squares~(-0.38). The ${\pm}0.4$ variations in $\epsilon$ mostly act to
reduce the amplitude of the scaling relation, while making the slope
slightly shallower with $\Delta p{=}-0.03$ and $-0.05$, respectively.

To answer the question raised at the beginning of this subsection,
Figure~\ref{fig:slope} demonstrates that for galaxies in the \eagle{}
simulation, the slope in the mass-metallicity relation at fixed $\sfr$ is
closely linked to $\alpha{-}\beta$, while the slope in the SFR-metallicity
relation at fixed $\ms$ is directly tied to $\beta$.  Such a tight
connection between the two sets of slopes suggests that, despite the
galaxies are generally out of equilibrium, our comprehensive NE-CEM could
still provide a promising avenue to extracting the underlying physics of
galactic outflows from the observed $\zgas(\msobs,\sfrobs)$ relation.

\subsection{Constraining mass loading from gas-phase metallicities: a mock
test with \eagle}
\label{subsec:mock}

To test the feasibility of constraining mass loading from metallicities, we
perform a Bayesian inference analysis over the \eagle{} simulation data
using our comprehensive NE-CEM described in \S\ref{subsec:assembly}.  We
employ a mock data set of the metallicity-stellar mass-SFR relation at
$z{=}0.1$, using the mean oxygen abundances $\zo^{0.1}$ of the same 28 bins
of ($\ms^{0.1}, \sfr^{0.1}$) as in Figure~\ref{fig:sfheagleall} and
\ref{fig:ceheagleall}. In addition, we convert the values of $\zo$ to
$\zgas{\equiv} 12{+}\lg(\mathrm{O/H})$ using $\zgas{=}12+\lg(\zo/(16
X_{\mathrm{ISM}}))$, where $X_{\mathrm{ISM}}$ is the hydrogen mass fraction
in the ISM and we adopt a constant $X_{\mathrm{ISM}}{=}0.7$.
Therefore, the data vector $\bm{x}$ comprises of 28
elements and we adopt the standard errors of the mean as the mock
uncertainties of $\zgas$, i.e., the diagonal errors of the
uncertainty matrix $\textbfss{C}$. We do not consider covariance between
different bins and set all the off-diagonal terms of $\textbfss{C}$ to
zero.

We predict the oxygen abundances for the 28 bins of ($\ms^{0.1},
\sfr^{0.1}$) as our model vector $\bm{\bar{x}}$ using the NE-CEM with nine
model parameters $\bm{\theta}{\equiv}\{A_\tau,B_\tau,A_0,B_0,\epsilon,M_{g,0},f,\alpha,\beta\}$.
We assume a Gaussian likelihood model
to compute the likelihood distribution of $\bm{x}$ given $\bm{\theta}$
\begin{equation}
    p(\bm{x} | \bm{\theta}) \propto \exp \left\{ - \frac{1}{2} \left(\bm{x}
    - \bm{\bar{x}(\theta)}\right)^T \textbfss{C}^{-1} \left(\bm{x} -
    \bm{\bar{x}(\theta)}\right)\right\}.
\end{equation}
For the prior distributions, we apply Gaussian priors informed by the
\eagle{} measurements on $A_\tau{\sim}\mathcal{N}(-1.11, 0.1^2)$ and
$A_0{\sim}\mathcal{N}(-0.52, 0.05^2)$, respectively, while adopting flat,
uninformative priors on the other parameters.

\begin{figure*}
\centering\includegraphics[width=0.8\textwidth]{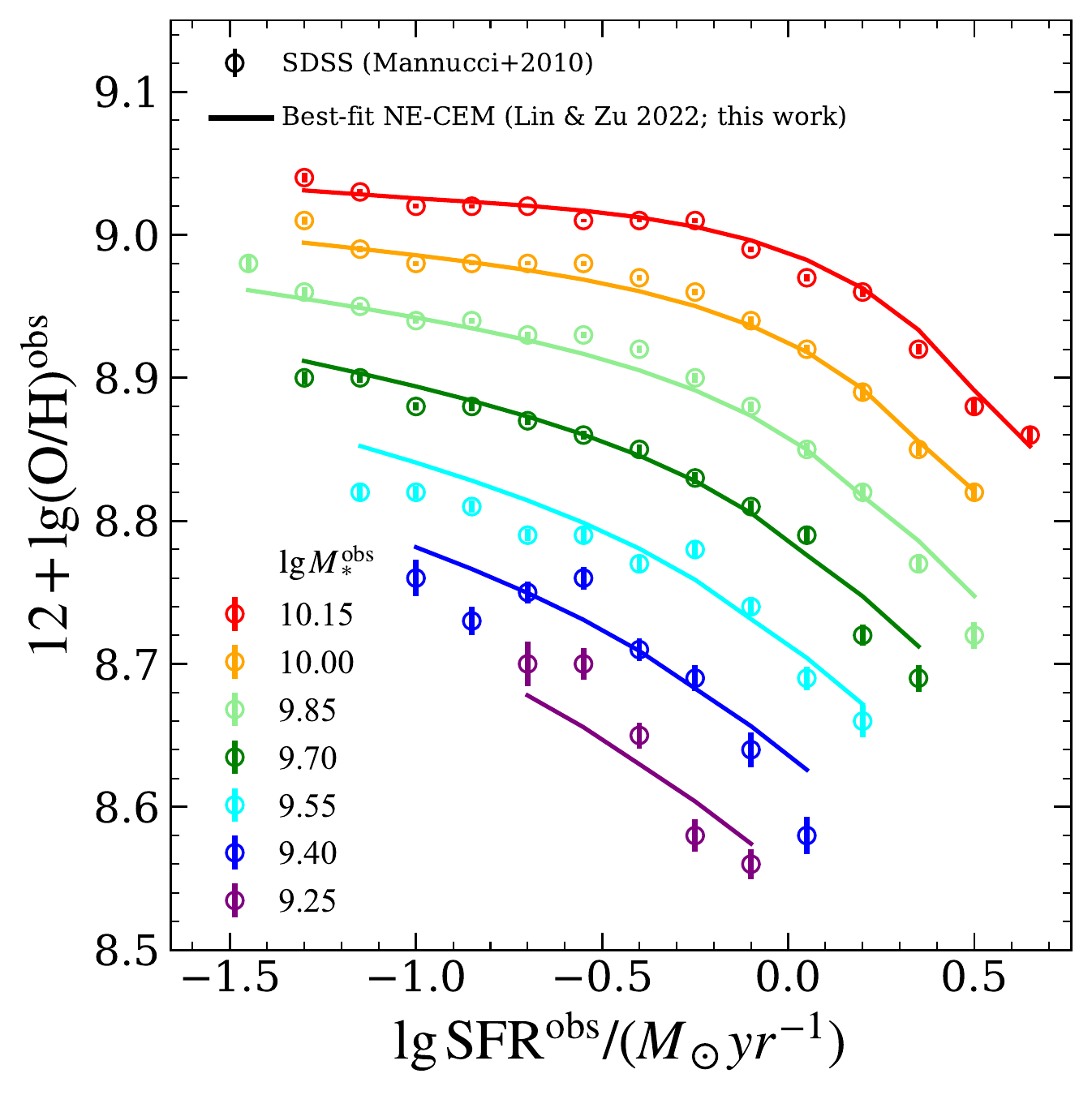}
    \caption{Comparison between the $\zgas(\msobs,\sfrobs)$ relation
    measured by \citet{Mannucci2010} from SDSS~(circles with errorbars) and
    that predicted by our posterior mean NE-CEM~(curves). Different colours
    correspond to galaxies in different $\msobs$ bins, indicated by the
    legend in the bottom left corner. NE-CEM provides an excellent
    description of the measurements from SDSS.
     \label{fig:obs_sdss}}
\end{figure*}

Given the data vector $\bm{x}$, model vector $\bm{\bar{x}}$, and the error
matrix $\textbfss{C}$, the posterior probability is proportional to the
product of the likelihood and the prior probability $p(\bm{\theta})$
\begin{equation}
    p(\bm{\theta} \mid \bm{x}) \propto p(\bm{x} | \bm{\theta}) \, p(\bm{\theta}).
\end{equation}
For the model inference, we employ the affine invariant Markov Chain Monte
Carlo~(MCMC) ensemble sampler \texttt{emcee}~\citep{Foreman-Mackey2013}.
We run the MCMC sampler for $3200000$ steps for our analysis to ensure its
convergence, and derive the posterior constraints after a burn-in period of
$240000$ steps.

The parameter constraints from our mock test are shown in
Figure~\ref{fig:mcmc_eagle}.  The histograms in the diagonal panels show
the 1D marginalised posterior distributions of each of the nine parameters,
and the contours in the off-diagonal panels are the $68\%$ and $95\%$ confidence regions
for each of the parameter pairs.  In the top right corners, we provide a
brief description of each model parameter using the relevant equations. The
median values and the 68 per cent confidence limits of the 1D posterior
constraints are listed on top of each histogram. In each diagonal panel,
solid and dashed lines indicate the posterior mean from the Bayesian
analysis and the best-fitting value from our calibrations against the
\eagle{} simulation , respectively. The two sets of best-fits are generally
in very good agreement with each other, except for $f$~($0.49\pm0.02$ vs.
${0.45}\pm{0.01}$) and $\alpha$~($-0.12\pm0.02$ vs. ${-0.15}\pm{0.02}$).
Naively, one may regard the apparent discrepancies in $f$ and $\alpha$
alarming, but applying the posterior means of $f{=}0.49$, $\alpha{=}{-}0.12$,
and $\beta{=}0.32$ to Equation~\ref{eqn:etacehmodel} actually provide a
reasonably good fit to the data points in Figure~\ref{fig:etafit}, yielding
a scatter~(0.08 dex) that is only slightly larger than obtained by the
direct calibration~(0.05 dex). Therefore, our Bayesian inference using the
metallicities at $z{=}0.1$ successfully recover the input parameters that
we directly measured from the histories of galaxies in the \eagle{}
simulation.

Finally, Figure~\ref{fig:obs_eagle} compares the $\zgas(\ms, \sfr)$ measured
at $z{=}0.1$~(circles with errorbars) with that predicted by the posterior
mean model~(lines with shaded uncertainty bands). Red, green, blue, and
purple indicate galaxies with $\lg\ms{=}10$, $9.75$, $9.50$, and $0.25$,
respectively. The posterior mean prediction provides excellent match to the
direct measurements from the \eagle{} simulation.

\section{A first-cut application to the SDSS data: understanding
FMR and galactic outflows}
\label{sec:sdss}

The mock test in \S\ref{subsec:mock} demonstrates that our nine-parameter
NE-CEM is capable of robustly recovering a comprehensive suite of galactic
histories, including star formation, chemical enrichment, and more
importantly, mass loading, from the $\zgas(\msobs, \sfrobs)$ relation
observed at a fixed epoch.  In this Section, we will apply the NE-CEM
analysis directly to the SDSS data, in hopes of gaining insights on the
nature of FMR and galactic outflows.  We emphasize again that a full
application of NE-CEM would include a joint SED-fitting of the galaxy
spectra, whereas the analysis below is a first-cut application that
demonstrates the efficacy of our model in interpreting real observations.

\subsection{Inferring mass loading from galaxy metallicities in SDSS}
\label{subsec:sdssanalysis}

\begin{figure*}
\centering \includegraphics[width=0.96\textwidth]{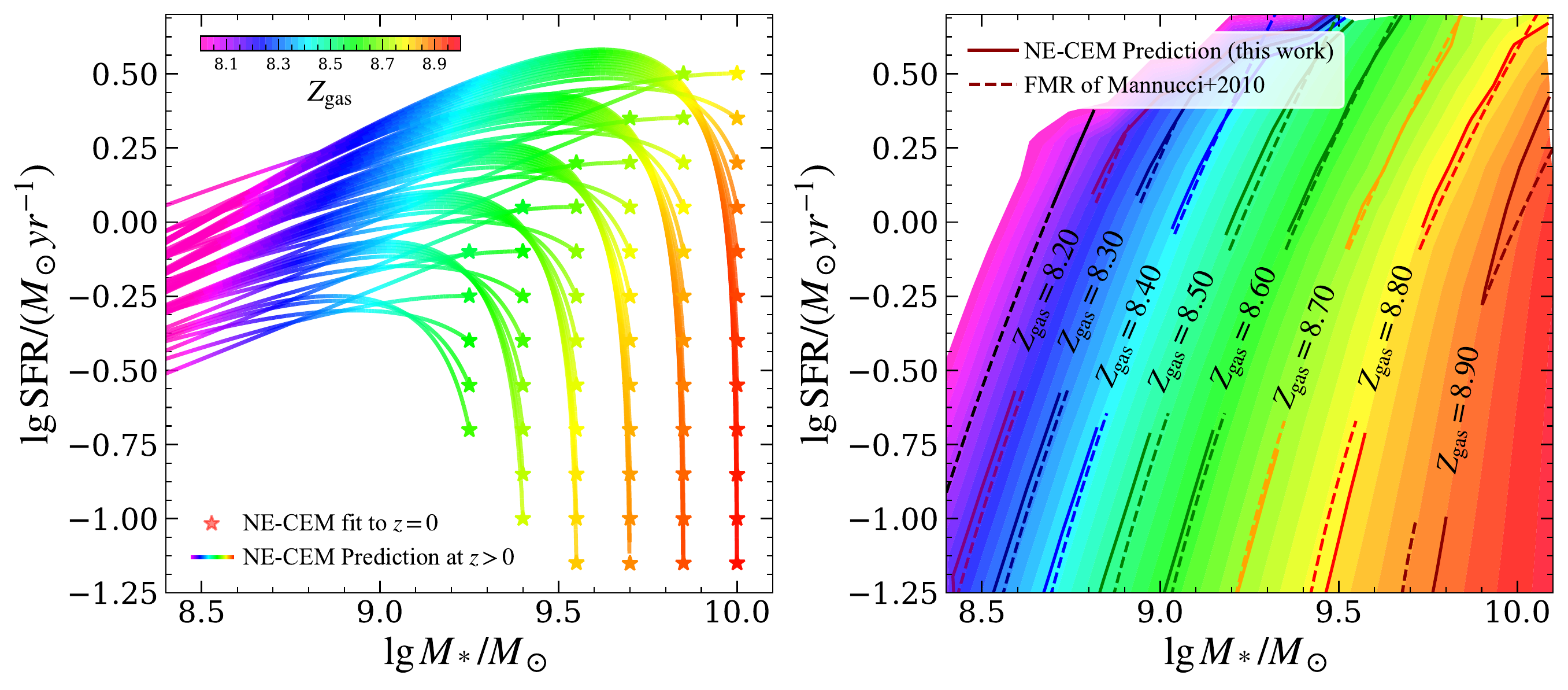}
    \caption{The emergence of a ``fundamental metallicity relation''~(right
    panel) from the coherent metal enrichment of different galaxies along
    their trajectories on the phase diagram of $\ms$ vs. $\sfr$~(left
    panel), predicted by our posterior mean NE-CEM.  The rainbow colours in
    both panels indicate the level of metal enrichment, ranging from
    $\zgas{=}8.2$~(purple) to $9.0$~(red) as coded by the colourbar in the
    top left corner of the left panel. {\it Left panel:} Each individual
    rainbow-coloured curve corresponds to one of the 76 bins of ($\msobs$,
    $\sfrobs$)~(i.e., an average galaxy) in Figure~\ref{fig:obs_sdss},
    delineating the chemical enrichment of the average galaxy as it evolves
    from low to high $\ms$ before arriving at the final mass of $\msobs$
    observed by SDSS at $z{=}0$~(star symbols). {\it Right panel:} Colour
    contours are the 2D metallicity map of galaxies on the $\ms$ vs. $\sfr$
    plane, predicted by the posterior mean NE-CEM by populating the phase
    diagram on the left panel with trajectories of a dense grid of average
    galaxies. Solid and dashed contour lines represent the FMR predicted
    from the NE-CEM and that proposed by \citet{Mannucci2010},
    respectively. \label{fig:fmr_sdss} }
\end{figure*}

We adopt the $\zgas(\msobs, \sfrobs)$ measurements listed in the table 1.
of \citet{Mannucci2010} as our input data. In particular, We employ seven
stellar mass bins centered at $\lg\msobs{=}9.25$, $9.40$, $9.55$, $9.70$,
$9.85$, $10$, and $10.15$, leaving out the higher mass galaxies that are
more likely dominated by AGN feedbacks.  For the individual ($\msobs$,
$\sfrobs$) bins, we only select those with galaxy number counts larger than
100 to ensure robust average metallicity measurements. For the $\sfrobs$
bins at $\lg\ms{=}9.25$, we reduce the number count threshold to 70 as the
overall number of low-mass galaxies is low. In total, we employ average
metallicity measurements for 76 bins in ($\msobs$, $\sfrobs$) as the data
vector, and adopt the errors on the mean as the measurement uncertainties.

Following the same methodology as in \S\ref{subsec:mock}, we apply our
NE-CEM to the 76 data points of the $\zgas(\msobs, \sfrobs)$ relation.  For
each bin of fixed $\msobs$ and $\sfrobs$, we start evolving the NE-CEM from
$\zo{=}10^{-4}$~(i.e., $\zgas{=}6.92$) at the epoch when the galaxies have
accumulated 1\% of their final mass. We have
tested that the final state of those galaxies does not vary if we push the
starting points to 0.1\% of their mass or reduce
the initial metallicity by another factor of ten, as the CEHs quickly
settle into the correct enrichment trajectories after the galaxies start
evolving.

The parameter constraints from SDSS are shown in
Figure~\ref{fig:mcmc_sdss}, with exactly the same format as
Figure~\ref{fig:mcmc_eagle}. All the parameters are generally well
constrained by the SDSS data, but with some prominent differences compared
to the \eagle{} predictions. Firstly, the constraint on $\epsilon$ is
$1.45{\pm}0.02$, while in \eagle{} it is $0.91{\pm}0.08$. As pointed out in
\S\ref{subsec:feature}, the difference caused by $\epsilon{>}1$ vs. ${<}1$
in the CEHs between SDSS and \eagle{} galaxies is small, although the two
could have distinct paths of enrichment in the future. Secondly, based on
the constraints on $\{A_\tau$, $B_\tau$, $A_0$, $B_0\}$, the SDSS data
require the galaxies to have rather different SFHs than predicted by
\eagle{}~(e.g., the posterior distribution of $A_{\tau}$ is shifted by 0.32
from the prior), so that younger galaxies started out much later than their
older counterparts with the same observed stellar mass. This discrepancy
between the \eagle{} and SDSS results is largely induced by the differences
in the $\ssfr$ between the \eagle{} predictions and SDSS
observations~\citep{Schaye2015}.  Lastly, the SDSS data require $\lg\eta$
to scale more strongly with $\ms$~($\alpha{=}{-}0.222{\pm}0.004$) than in
\eagle{}~(${-}0.12{\pm}0.01$), but have significantly weaker dependence on
$\ssfr$~($\beta{=}0.078{\pm}0.003$) than \eagle{}~($0.32{\pm}0.07$).

A more visually appealing presentation of our constraints is displayed in
Figure~\ref{fig:eh_sdss}, where we show the CEHs, MLHs, and SFHs predicted by the
posterior mean NE-CEM in the main and two inset panels, respectively~(top row
and the first three rows in the bottom, with $\msobs$ increasing from left
to right, top to bottom). The shaded bands are the $1{-}\sigma$
uncertainties, and the dashed portion of the SFHs indicates the histories
before the galaxies have gained 1\% of their final mass.
The SFHs exhibit a clear ``downsizing'', i.e., the more massive galaxies
started forming stars and enrich
their ISM earlier than the less massive
ones~\citep{Spitoni2020}.  Interestingly, for galaxies with different
$\sfrobs$ in the same $\msobs$ bin, although the SFHs have drastically
different shapes, their star formation~(inset panel in the bottom right)
conspires with outflows~(inset panel in the top left) to produce very
coherent CEHs~(main panel). In addition, the equilibrium timescales are
indicated by the short vertical lines underneath the top x-axis of each
panel should they occur before $z{=}0.1$ --- the more massive galaxies are
progressively closer to equilibrium, and at the same mass galaxies with
lower $\sfrobs$ are preferentially closer to equilibrium and have older
age~\citep{DuartePuertas2022}.  The bottom rightmost panel shows the
excellent agreement between the MZR measured by averaging the observed
$\zgas(\msobs, \sfrobs)$ at each $\msobs$~(circles with errorbars) and that
predicted by the posterior mean NE-CEM~(solid curve; not a fit).

Finally, Figure~\ref{fig:obs_sdss} shows the comparison between the
$\zgas(\msobs, \sfrobs)$ relation measured by \citet{Mannucci2010}~(circles
with errorbars) and predicted by the posterior mean NE-CEM~(curves with
uncertainty bands), with the colours indicating the observed stellar
mass~(increasing from purple to red). Overall, the NE-CEM provides
excellent fits to the SDSS data points.  In particular, the agreement in
the high-mass bins, where the observed metallicities exhibit a plateau at
the low-$\sfr$ end but steeply declines at the high-$\sfr$ end, is highly
nontrivial.  In the posterior mean NE-CEM, the shallower slope at the
low-$\sfr$ end is caused by the relatively short equilibrium timescale of
the massive, almost-quenched galaxies, which have already arrived at the
equilibrium metallicities by the time they are observed.

\subsection{Can we reproduce the FMR using our non-equilibrium CEM?}
\label{subsec:fmr}

The posterior mean NE-CEM obtained in \S\ref{subsec:sdssanalysis} not only
provides an excellent description of the $\zgas(\msobs, \sfrobs)$ relation
of galaxies observed at $z{\sim}0$, but also predicts the entire histories
of chemical enrichment for those galaxies, i.e., the $\zgas(\ms, \sfr | z)$
relation of their progenitors in the past~($z{>}0$). For any successful
model of CEM, it is imperative to explain the apparent invariance of the
$\zgas(\msobs, \sfrobs)$ relation with redshift as suggested by the
observations~\citep[e.g., ][]{Mannucci2010}. With most of the
galaxies out of equilibrium, it would be interesting to find out if an FMR
still manage to emerge from the progenitor galaxies at $z{>}0$.

Figure~\ref{fig:fmr_sdss} explores the chemical enrichment of galaxies on
the star-formation ``phase'' diagram~(i.e., 2D plane of $\ms$ vs. $\sfr$)
predicted by the posterior mean NE-CEM~(left panel), as well as the
resulting 2D map of $\zgas$ on the same diagram~(right panel).  In the left
panel, the rainbow-coloured curves indicate the chemical enrichment~(colour
gradient) along the star-forming trajectories~(curves) of each of the 76
bins of galaxies on the phase diagram, colour-coded by the colourbar on the
top left. The star symbols indicate the end points of the trajectories at
$z{=}0$, which correspond to the curves shown in Figure~\ref{fig:obs_sdss}
for the six stellar mass bins. As expected, the six bundle of trajectories
are self-similar, because galaxies of different masses follow the same
\powexp{} family of SFHs. In particular, the trajectory of a more massive
galaxy can be obtained by shifting that of a less massive galaxy diagonally
to the top right of the phase diagram, i.e., via the rescaling of
$\dot{M}_{*,0}$ in Equation~\ref{eqn:sfh}; Within the same bundle~(i.e.,
the same $\msobs$), the trajectories of galaxies with different $\sfrobs$
are also approximately self-similar by the rescaling of the time variable
by their respective $\tausfh$ in Equation~\ref{eqn:sfh}.

More important, the chemical enrichment along different but self-similar
star-forming trajectories appear coherent on the $\ms$ vs. $\sfr$ diagram
in the left panel. This coherence can be qualitatively understood as
follows.  Take the six most quiescent ``galaxies''~(i.e., the lowest
$\sfrobs$ curves of each of the six $\msobs$ bins) for an example, the
trajectory of any of the five more massive galaxies~(e.g., galaxy B) is
approximately a rescaled version of that of the least massive
galaxy~(galaxy A) by
$\mathcal{R}{=}\dot{M}_{*,0}^{B}/\dot{M}_{*,0}^{A}{>}1$, which is
equivalent to shifting the trajectory diagonally by $\lg \mathcal{R}$ on
the log-log diagram. If chemical enrichment is independent of $\mathcal{R}$
and thus retains the perfect self-similarity, the loci of constant $\zgas$
would also appear diagonally on the diagram, producing an FMR in the form
of $\zgas{\equiv}\zgas(\mu_{x}{=}\lg\ms {-} x\lg\sfr)$ with $x{=}1$.
However, the chemical enrichment in NE-CEM is governed by
Equation~\ref{eqn:dzodtpowexp}, which depends on $\mathcal{R}$ via the
dependence of $\eta$ on $\ms$ and $\sfr$ in Equation~\ref{eqn:etacehmodel}.
In particular, for two loci ($\ms$, $\sfr$) and ($\mathcal{R}\ms$,
$\mathcal{R}\sfr$) on the trajectories of A and B, respectively, the
enrichment rate $\dzodt$ will be faster at ($\mathcal{R}\ms$,
$\mathcal{R}\sfr$) for galaxy B than at ($\ms$, $\sfr$) for galaxy A, due
to the smaller mass-loading factor of outflows in the more massive
galaxy~(as $\eta_{B}{=}\eta_{A}^{\mathcal{R}^\alpha}{<}\eta_{A}$ for
$\alpha{<}0$).  Therefore, the chemical enrichment will be lagging behind
the self-similarity prediction, thereby making $x<1$ while staying coherent
on the phase diagram.

\begin{figure}
\centering\includegraphics[width=0.48\textwidth]{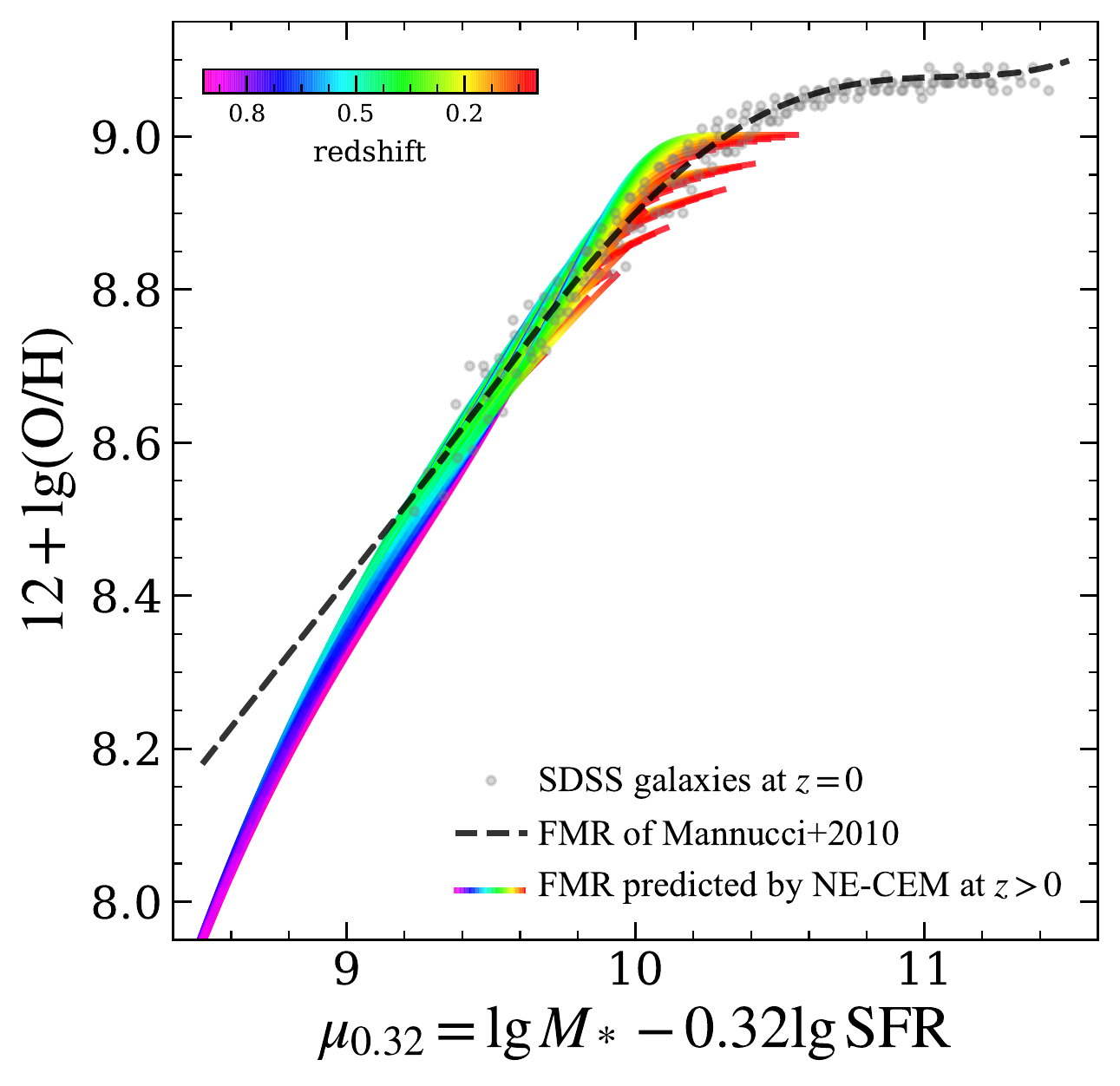}
\caption{Comparison between the FMR
    predicted by our posterior mean NE-CEM~(coloured curves) and that
    proposed by \citet{Mannucci2010}~(dashed curve). Each rainbow-coloured
    curve is the predicted track of an average galaxy from one of the 76 bins in
    Figure~\ref{fig:obs_sdss}, colour-coded by the redshifts according to
    the colourbar on the top left. Gray dots represent the individual
    measurements from SDSS at $z{=}0$ by \citet{Mannucci2010}, including
    massive galaxies~($\msobs{>}10^{10.2}\msol$) dominated by AGN
    feedbacks.
 \label{fig:mu_z_sdss} }
\end{figure}

To find out whether the coherence seen in the left panel of
Figure~\ref{fig:fmr_sdss} would indeed produce the observed FMR, we employ
our posterior mean NE-CEM to predict the SFHs and CEHs for mock galaxies on
a dense grid of $\msobs$ and $\sfrobs$, in order to have a full coverage of
the phase diagram.  We then compute the average $\zgas$ in each state of
$\ms$ and $\sfr$ on the diagram, shown by the 2D map of $\zgas$ on the
right panel of Figure~\ref{fig:fmr_sdss} and colour-coded by the same
colourbar in the left panel. The solid and dashed contour lines indicate
the iso-metallicity contours predicted by the posterior mean NE-CEM and the
FMR inferred by \citet{Mannucci2010}~(their equation 4 with $x{=}0.32$),
respectively.  The two sets of contour lines are largely aligned and
overlapping, exhibiting remarkable consistency between the NE-CEM
prediction and the observed FMR for all galaxies at $z{\geq}0$. Such
consistency also confirms our hypothesis in the Introduction that an FMR
with the correct $x{=}0.32$ could emerge out of the coherent histories of
metal enrichment between different star-forming galaxies, without the need
to impose chemical equilibrium.

\begin{figure*}
\centering\includegraphics[width=0.8\textwidth]{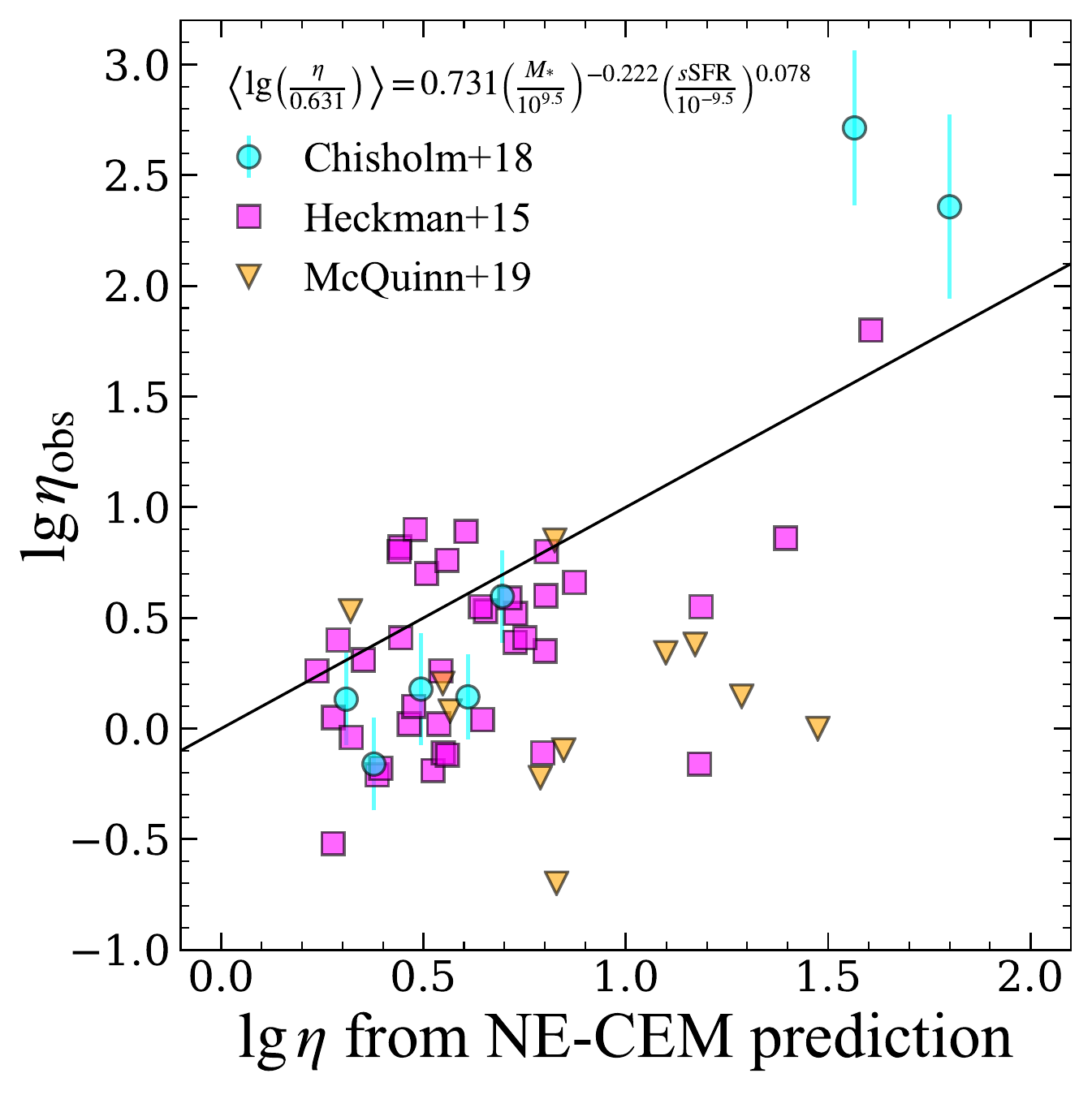}
    \caption{Comparison between the observed mass loading factors~(y-axis)
    and the predictions by our posterior mean NE-CEM from the observed
    $\ms$ and $\ssfr$ of those galaxies~(x-axis) using
    Equation~\ref{eqn:etabest}~(also shown in the top left corner).  Cyan
    circles with errorbars, magenta squares, and orange triangles are the
    measurements from~\citet{Chisholm2018}, \citet{Heckman2015},
    and~\citet{Mcquinn2019}, respectively, with the solid straight line
    indicating the one-to-one relation.
	\label{fig:eta_2d}}
\end{figure*}

\begin{figure*}
\centering\includegraphics[width=0.8\textwidth]{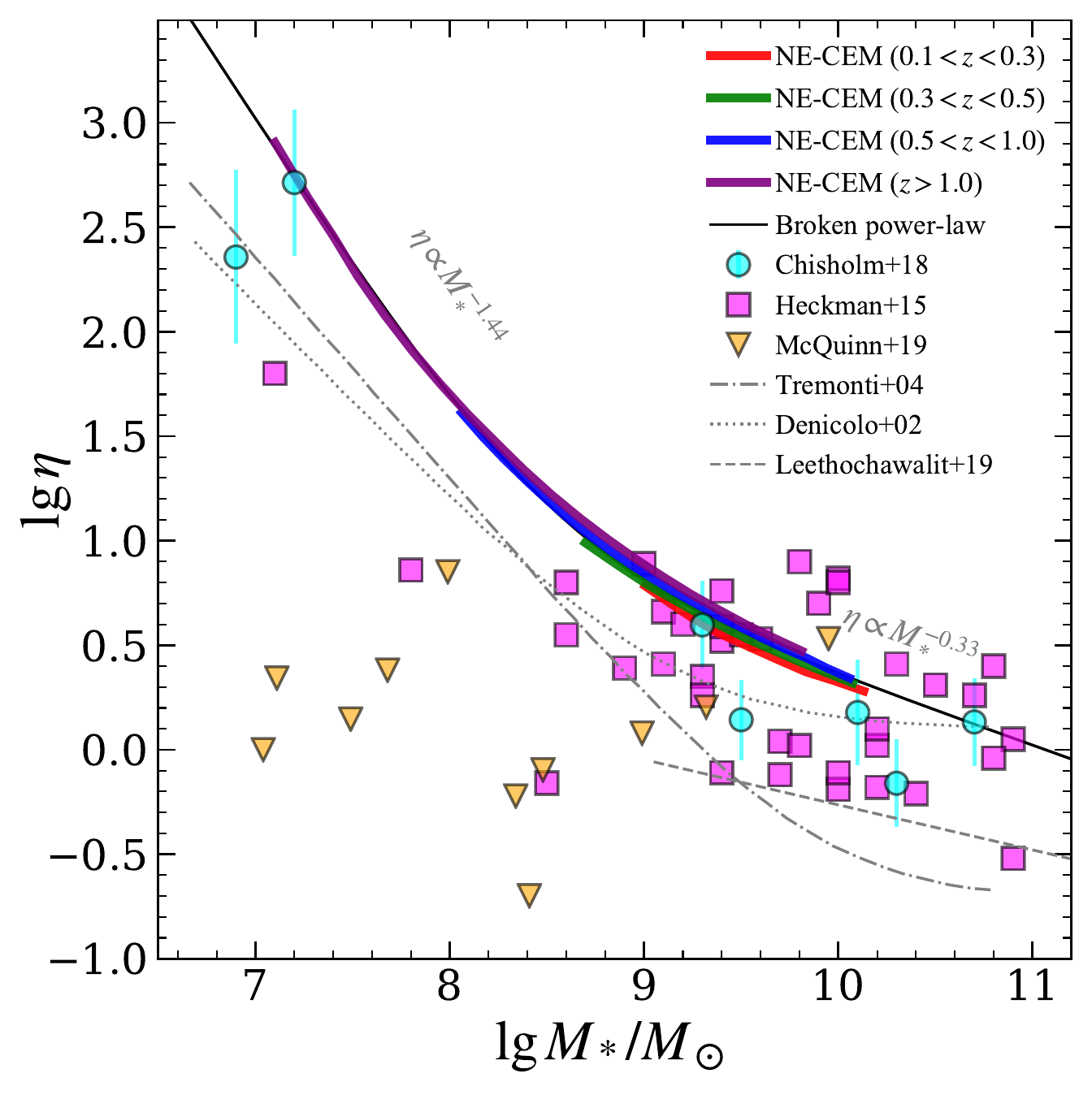}
    \caption{Comparison between the stellar mass-dependence of the mass
    loading factor inferred from our NE-CEM constraint~(thick solid curves)
    and that from observational measurements~(symbols) as well as other
    empirical constraints~(gray curves). We convert our constraint in
    Equation~\ref{eqn:etabest} to an dependence of $\eta$ on $\ms$ at
    different $z$ by marginalising over the dependence on $\ssfr$, yielding
    four $\zgas$-$\ms$ relations at $0.1{<}z{<}0.3$~(thick red curve),
    $0.3{<}z{<}0.5$~(thick green), $0.5{<}z{<}1.0$~(thick blue), and
    $z{>}1.0$~(thick purple). Solid black curve is a broken power-law
    fit~(Equation~\ref{eqn:etamass}) to the four relations, with
    $\eta\propto\ms^{-1.1}$ and $\eta\propto\ms^{-0.33}$ at the low- and
    high-mass ends, respectively.  Cyan circles with errorbars, magenta
    squares, and orange triangles indicate the direct measurements~(same as
    in Figure~\ref{fig:eta_2d}) from~\citet{Chisholm2018},
    \citet{Heckman2015}, and~\citet{Mcquinn2019}, respectively.  Dashed
    curve represents the empirical constraint from
    \citet{Leethochawalit2019}, whereas dotted and dot-dashed curves
    indicate the $\eta$-$\ms$ relations inferred by \citet{Peeples2011}
    using the MZR measured by \citet{Denicolo2002} and
    \citet{Tremonti2004}, respectively.  \label{fig:eta_sdss} }
\end{figure*}

Alternatively, Figure~\ref{fig:mu_z_sdss} shows the enrichment tracks on
the $\zgas$ vs. $\mu_{0.32}{\equiv}\lg\ms{-}0.32\lg\sfr$ plane of the 76
bins of galaxies, predicted by our posterior mean NE-CEM~(rainbow-coloured
curves), colour-coded by redshift according to the colourbar on the top
left. Gray dots are the data points from \citet{Mannucci2010}~(but
including those with $\msobs>10^{10.2}\msol$), while the black dashed curve
indicates the best-fitting FMR by \citet{Mannucci2010}. Our predicted
tracks closely follow \citeauthor{Mannucci2010}'s best-fit at
$\mu_{0.32}{>}9$, but drop more steeply with decreasing $\mu_{0.32}$ below
$\mu_{0.32}{=}9$. Future metallicity observations for a large sample of
dwarf galaxies at high redshifts could help test our predicted deviation
from the SDSS extrapolation from $L_*$ galaxies.  In addition, the scatter
between different predicted tracks is consistent with the small scatter in
the SDSS FMR inferred by~\citet{Mannucci2010}~(0.05 dex).  This consistency
is not only a powerful validation of the success of our NE-CEM framework,
but also naturally explains the origin of the tightness of the FMR ---
star-forming galaxies in the Universe make stars, drive winds, and enrich
gas in non-equilibrium yet strikingly coherent fashion.

To answer the question raised by the title of this subsection, the chemical
enrichment of galaxies predicted by our NE-CEM at $z{>}0$ exhibit excellent
consistency with the $z{=}0$ FMR proposed by~\citet{Mannucci2010}, thereby
successfully reproducing the redshift invariance of the FMR without
resorting to equilibrium.  We emphasize again that the rainbow curves and
contours in Figure~\ref{fig:fmr_sdss} and Figure~\ref{fig:mu_z_sdss}) are
{\it not} direct fits to the FMR proposed by \citet{Mannucci2010}, but
enrichment trajectories {\it predicted} by the NE-CEM at $z{>}0$. This is
non-trivial because unlike the equilibrium models, the progenitors of the
current-day galaxies at $z>0$ may not obey the same $\zgas(\msobs,
\sfrobs)$ relation observed at $z{=}0$ when assuming non-equilibrium.

\subsection{Physical implication of our constraints on $\eta$ and comparison with
direct observations}
\label{subsec:winds}

Our posterior mean mass-loading factor of galactic outflows~(assuming
entirely entrained ISM) is\footnote{This is equivalent to a constraint on
the metal-loading factor $\zeta$ if the outflows are over-enriched with ISM
entrainment fraction below unity. See Equation~\ref{eqn:metalloading}.}
\begin{equation}
\lg \left(\frac{\eta}{0.631}\right) = 0.731
    \left(\frac{\ms}{10^{9.5}\msol}\right)^{-0.222}
    \left(\frac{\ssfr}{10^{-9.5}yr^{-1}}\right)^{0.078}.
    \label{eqn:etabest}
\end{equation}
While the anti-correlation between $\eta$ and $\ms$ is generally expected
by various galaxy formation models, the dependence of $\eta$ on $\ssfr$ or
$\sfr$ is still under debate.  For instance, using a high-resolution
hydrodynamical simulation of an isolated magnetised Milky Way-like disc
galaxy, \citet{Wibking2022} obtained a positive linear correlation between
$\eta$ and $\sfr$, and found that the predicted $\eta$ is indistinguishable
from a non-magnetic simulation if the SFH is controlled. However, using a
suite of hydrodynamic cosmological zoom-in simulations with an explicit
stellar feedback model, \citet{Muratov2015} predicted that $\eta$ is
independent of $\sfr$. More intriguingly, using a suite of
parsec-resolution local galactic disk simulations with explicit modelling
of the ISM, \citet{Kim2020} predicted that the mass-loading factors of the
cool gas outflows decrease steeply with the surface density of star
formation of different model galaxies, albeit with strong variation in the
initial gas surface density.

We detect a weakly positive but statistically significant correlation
between $\eta$ and $\ssfr$~($\beta{=}0.078{\pm}0.003$). To further test the
necessity of having a non-zero $\beta$ in our mass-loading model, we repeat
the MCMC analysis of \S\ref{subsec:sdssanalysis} while fixing the value of
$\beta$ to be zero, and find that such a model is incapable of reproducing
the observed $\zgas(\msobs, \sfrobs)$ relation at $z{=}0$. Therefore, our
NE-CEM analysis makes a strong observational argument for the existence of
a positive correlation between $\eta$ and $\ssfr$ at fixed $\ms$.

Figure~\ref{fig:eta_2d} compares the observed mass-loading factors of three
sample of galaxies with the predictions by our posterior mean
NE-CEM~(Equation~\ref{eqn:etabest}).  From the outflowing warm ionized gas
traced by ultraviolet~(UV) metal absorption lines in low-redshift starburst
galaxies~($\ms{\sim}10^{7.5}{-}10^{11}\msol$), \citet{Heckman2015} found
the galactic winds travelling at velocities ${\sim}100{-}500\,\kms$ with an
average total column density around $10^{21}\mathrm{cm}^{-2}$. By assuming
isotropic winds at twice the starburst radii, they estimated the
mass-loading factors to be $\eta{\sim}1{-}10$, with weak anti-correlations
with $\ms$ or $\sfr$. However, our predictions of $\eta$ from the observed
$\ms$ and $\sfr$ are in excellent agreement with the measurements
by~\citet[magenta squares]{Heckman2015}.

Using similar UV observations but very different modelling assumptions,
\citet{Chisholm2018} found that the outflows are highly enriched compared
to the ISM and the metal-loading factors exhibit strong anti-correction
with $\ms$~(cyan circles with errorbars).  From H$\alpha$ narrowband deep
imaging, \citet{Mcquinn2019} estimated the mass-loading factors to be
$0.2{-}7$ for a dozen near-by dwarf galaxies with
$\ms{\sim}10^7{-}10^{9.3}\msol$, but found little dependence of $\eta$ on
$\ms$ or $\sfr$~(orange triangles). Our NE-CEM constraint agrees reasonably
well with both the \citeauthor{Chisholm2018} and \citeauthor{Mcquinn2019}
measurements in the $\eta{<}10$ regime~(i.e., galaxies with
$\ms>10^9\msol$) In the high-$\eta$ regime occupied by the dwarf galaxies,
our constraint slightly under-predicts $\eta$ for the two dwarf galaxies in
\citeauthor{Chisholm2018}, but over-predicts $\eta$ for the four low-mass
galaxies in \citeauthor{Mcquinn2019}. However, since the two sets of
measurements are not consistent with each other for the low-mass systems,
more observations of the outflows in the dwarf galaxies are needed to test
the predictions from our NE-CEM constraint.

To facilitate the comparison with other empirical constraints of the
dependence of mass-loading factors on stellar mass, we convert our
constraint on $\eta$ as in Equation~\ref{eqn:etabest}~(i.e., as a function
of $\ms$ and $\ssfr$) to that on the relations between $\eta$ and $\ms$ at
different redshifts, using the SFHs predicted by the posterior mean NE-CEM.
The result is shown in Figure~\ref{fig:eta_sdss}, where we compare the
stellar mass dependences of the $\eta$ predicted by our posterior mean
NE-CEM at $0.1{<}z{<}0.3$~(thick red curve), $0.3{<}z{<}0.5$~(thick green),
$0.5{<}z{<}1.0$~(thick blue), and $z{>}1$~(thick purple) with a suite of
empirical constraints~(thin curves) and observational results~(symbols;
same as those in Figure~\ref{fig:eta_2d})
from the literature.  The four coloured curves exhibit a steeper trend with
$\ms$ at the low mass than at the high mass end, which can be described by
a broken power-law that scales as $\ms^{-1.44}$ and $\ms^{-0.33}$ at the
low and high-mass ends, respectively.  In particular, the solid black curve
in Figure~\ref{fig:eta_sdss} is
\begin{equation}
    \eta = 7
    \left(1 + \frac{\ms}{10^{8.5}\msol}\right)^{1.10}
    \left(\frac{\ms}{10^{8.5}\msol}\right)^{-1.44}.
    \label{eqn:etamass}
\end{equation}
Note that Equation~\ref{eqn:etamass} expresses $\eta$ as a function of
$\ms$ after marginalising over the dependences on $\ssfr$, whereas in our
full constraint~(Equation~\ref{eqn:etabest}) $\eta$ scales as
$\ms^{-0.222}$ at any given $\ssfr$.

As expected from Figure~\ref{fig:eta_2d}, our constraint on $\eta$~(thick
coloured curves) and its extrapolation to both higher and lower mass
ranges~(solid black curve) exhibit a broad agreement with the direct
measurements~(symbols). Our constraint is slightly higher than the
empirical constraints from \citet{Peeples2011} using the MZR derived in
\citet[][dot-dashed curve]{Tremonti2004} and \citet[][dotted
curve]{Denicolo2002}, as well as the constraint from the Mg abundances of
quiescent galaxies in massive clusters~\citep[][dashed
curve]{Leethochawalit2019}.  The discrepancy is probably due to the
different model assumptions on equilibrium and mass-loading, but some of it
can be at least partially explained by the different metallicity
calibrations in the data~\citep{Kewley2008}.

\section{Conclusion}
\label{sec:conc}

In this paper, we have developed a comprehensive framework of
non-equilibrium chemical evolution model~(NE-CEM) by explicitly tracking
the average star-formation history of galaxies and the mass-loading history
of stellar feedback-driven outflows along the SFH. After exploring the SFHs
of galaxies in the \eagle{} hydrodynamical simulation, we discover that a
simple yet flexible model~(\powexp{}) can accurately describe the SFHs of
the simulated galaxies, thereby allowing us to robustly reconstruct the
average SFHs for the observed galaxies at fixed $\msobs$ and $\sfrobs$.

To explore the parametrisation of the mass-loading factor $\eta$, we
firstly develop a novel method to reconstruct the evolution of $\eta$ from
the chemical enrichment history of galaxies in hydrodynamical simulations.
After applying the reconstruction method to the \eagle{} simulation, we
discover that the mass-loading at any epoch can be accurately~(with a
scatter of 0.05 dex) described by the stellar mass $\ms$ and specific star
formation rate $\ssfr$ of galaxies at that epoch, so that $\lg\eta\propto
\ms^{\alpha}\ssfr^{\beta}$~(with $\alpha{=}{-}0.15$ and $\beta{=}0.34$ in
\eagle{}). Encouragingly, our chemically-inferred mass-loading factors
exhibit remarkable agreement with those measured kinematically by
\citet{Mitchell2020} from tracking wind particles in the \eagle{}
simulation.  Such an agreement is highly nontrivial, and since the
\citeauthor{Mitchell2020} measurement is somewhat analogous to the
down-the-barrel measurements of $\eta$ in the real observations, the
agreement greatly reinforces our belief that we can potentially constrain
wind kinematics and energetics by applying our NE-CEM method to the real
data.

By tracking the chemical enrichment of ISM along the \powexp{} SFH with
time-dependent mass-loading of outflows, we can predict the
metallicity-stellar mass-SFR relation of the entire star-formation
main-sequence at any observed epoch, $\zgas(\msobs, \sfrobs)$, without the
need to assume some equilibrium.  The gas accretion history is implicitly
tracked via the modelling of the gas reservoir as
$\mgas\propto\sfr^{\epsilon}$. Despite the comprehensive predictive power,
our NE-CEM has only nine parameters in total.  We demonstrate the efficacy
of the NE-CEM framework in constraining the mass-loading properties of
galactic outflows, i.e., $\alpha$ and $\beta$, by performing extensive
analytic and mock tests using the $\zgas(\msobs, \sfrobs)$ measured from
\eagle{} at $z{=}0.1$.

As a first-cut application of our NE-CEM framework, we perform a Bayesian
inference analysis using the SDSS $\zgas(\msobs, \sfrobs)$ relation
measured by \citet{Mannucci2010}. The posterior mean NE-CEM not only
provides excellent description of the SDSS $\zgas(\msobs, \sfrobs)$
relation at $z{=}0$, but also correctly predicts the redshift invariance of
the so-called ``fundamental metallicity relation'', $\zgas(\mu_{0.32})$,
with most galaxies far from equilibrium. Therefore, equilibrium or a steady
gas reservoir is not a prerequisite for the existence of a fundamental
metallicity relation. In our NE-CEM framework, the $\zgas(\mu_{0.32}$
relation emerges out of the coherent histories of chemical evolution
between different star-forming galaxies, due to the self-similarity in
their star-forming and mass-loading behaviors.

We obtain a tight constraint on the mass-loading factors from SDSS as
\begin{equation}
\lg \left(\frac{\eta}{0.631}\right) = 0.731\pm0.002
    \left(\frac{\ms}{10^{9.5}}\right)^{-0.222\pm0.004}
    \left(\frac{\ssfr}{10^{-9.5}}\right)^{0.078\pm0.003},
    \label{eqn:etafinal}
\end{equation}
after marginalising over the nuisance parameters~(e.g., $\epsilon$) in the
NE-CEM.  This constraint is broadly consistent with various direct
observations and empirical constraints from the literature.  In particular,
using the observed stellar mass and SFRs of a sample of galaxies in the
local Universe, we predict their mass-loading factors and find good
agreement with the down-the-barrel
observations~\citep[e.g.,][]{Heckman2015, Chisholm2018}.  In addition, the
posterior mean NE-CEM predicts that the mass loading factor scales as
$\eta{\propto}\ms^{-1.44}$ and $\ms^{-0.33}$ at the low- and high-$\ms$
ends, respectively, after marginalising over the $\ssfr$ dependence. This
broken power-law behavior of $\eta$ is in reasonable agreement with other
empirical constraints from the mass-metallicity
relation~\citep{Peeples2011}.

Therefore, our constraint on $\eta$ provides an excellent benchmark for
different sub-grid models of stellar feedbacks in hydrodynamical
simulations and SAMs. Despite the increasing sophistication in the explicit
modelling of galactic winds in modern cosmological hydrodynamical
simulations~\citep{Khandai2015, Schaye2015, Pillepich2017, Hopkins2018,
Dave2019, Vogelsberger2020, Pakmor2022}, the effective mass-loading of the
simulated outflows on galactic scales should be roughly consistent with our
constraint, so as to reproduce the correct $\zgas(\msobs, \sfrobs)$
relations observed between $z{=}0$ and $z{=}2.5$.

Looking to the future, our method can be significantly improved in several
important aspects. The average SFHs can be more accurately inferred from
the stacked spectra of star-forming galaxies at fixed $\msobs$ and
$\sfrobs$, by adopting the \powexp{} SFH model~\citep[or some modified
variant; see e.g.,][]{Simha2014} during SED fitting. By the same token, the
average gaseous metallicities can be measured more robustly using the
direct method from auroral lines~\citep{Andrews2013} or self-consistently
from the overall SED fitting~\citep{Thorne2022}. With upcoming
spectroscopic surveys like the DESI~\citep{desi2022} and
PFS~\citep{Takada2014}, we expect our NE-CEM framework to provide a
promising avenue to unlocking the exquisite yet coherent histories of
chemical enrichment and stellar feedback in star-forming galaxies across
cosmic time.

\section*{Data availability}

The data underlying this article will be shared on reasonable request to the corresponding author.

\section*{Acknowledgements}

This article is dedicated to the memory of Dr. Yu Gao, from whom the
authors benefited tremendously discussing about the star formation law and
gas reservoirs at Xiamen University. We are indebted to David Weinberg for
his invaluable suggestions that have greatly improved the overall quality
of the paper.  We also thank the referee for the helpful comments and Junde
Chen for stimulating discussions at the early stage of this work. Y.L. and
Y.Z. acknowledge the support by the National Key Basic Research and
Development Program of China (No. 2018YFA0404504), the National Science
Foundation of China (12173024, 11621303, 11890692, 11873038), the science
research grants from the China Manned Space Project (No. CMS-CSST-2021-A01,
CMS-CSST-2021-A02, CMS-CSST-2021-B01), and the ``111'' project of the
Ministry of Education under grant No. B20019. Y.Z. acknowledges the
generous sponsorship from Yangyang Development Fund, and thanks Cathy Huang
for her hospitality during the pandemic at the Zhangjiang Hi-Technology
Park where he worked on this project.



\bibliographystyle{mnras}
\bibliography{reference.bib}
\bsp	
\label{lastpage}

\end{document}